\documentstyle[12pt,aaspp4]{article}

\def\ha{H$\alpha$}   
\def\noi{\noindent}

\begin{document}

\title{Radial Distribution of the Mass-to-Luminosity Ratio in 
Spiral Galaxies and Massive Dark Cores}
\author{Tsutomu TAKAMIYA \& Yoshiaki SOFUE}
\affil{Institute of Astronomy, University of Tokyo, Mitaka, 
Tokyo 181-8588, Japan} 

\begin{abstract} 
We derive radial profiles of the surface-mass-density for 19 spiral 
galaxies directly from their high-resolution rotation curves.
Using the corresponding luminosity profiles, we obtain the radial distribution 
of the mass-to-luminosity ratios ($M/L$) from the inner bulge ($\sim$ a 
few 100 pc) to the outer disk ($\geq$ 2$-$10 kpc) for 11 galaxies (with
inclination $<$ $70^{\circ}$ in order to reduce the influence 
of the interstellar extinction.  The $M/L$s in the bulges of two galaxies 
with sufficient resolution, NGC  4527 and NGC  6946, are found to increase 
steeply toward the center at radii $\sim$ 100$-$500 pc at rates of 
15$\pm$3 and 7$\pm$2 times per kpc, respectively.
Some other galaxies with fairly high resolution also show signs of an
increase toward the center. Such an increase may indicate the existence of 
a new component, a ``massive dark core'', which may be an object linking 
the bulge and a central black hole. 
Based on radial variations of the $M/L$, we further discuss the 
variation  of the dark-mass fraction in spiral galaxies. 
\end{abstract}


\section{Introduction}
A Radial variation of the mass-to-luminosity ratio in a galaxy 
(hereafter $M/L$) is a clue to investigate the distributions of 
visible and dark (invisible) masses (e.g., Kent 1986, 1987; 
Persic \& Salucci 1988, 1990; Salucci \& Frenk 1989; Forbes 1992; 
Persic et al. 1996; H\'{e}raudeau \& Simien 1997; Rubin et al. 
1997). However, most of the current studies have been devoted to 
investigate the dark halo, which is supposed to be closely related 
to galaxy formation and the stability of disks (e.g., Ostriker \& 
Peebles 1973; Athanassoula et al. 1987). On the other hand, there 
have been few systematic investigations of dark matter within a 
bulge and an inner disk, for which, however, high-resolution and 
high-quality rotation curves are required as well as corresponding 
luminosity profiles. 

Rotation curves (RCs) in disk and outer regions have been derived 
from position-velocity diagrams obtained by optical and HI-line 
spectroscopy (Rubin et al. 1980, 1982, 1985; Bosma 1981a; 
Mathewson et al. 1992; Persic \& Salucci 1995; Mathewson \& Ford 
1996; Clemens 1995; Sofue 1996, 1997; Honma \& Sofue 1997). 
Recently, we have obtained inner-most RCs by analyzing 
high-resolution CO, H$\alpha$, and [NII]-line spectroscopic data
(Sofue 1996, 1997; Sofue et al. 1997, 1998, 1999). Fig. 1 shows thus 
obtained RCs of Sb and Sc galaxies (e.g., Sofue 1999). 
All the Sb galaxies in our sample show very steep rises of rotation 
velocity in the central 100 $-$ 500 pc region. 
The Sc galaxies also show similar nuclear rises, except for a few 
small-mass galaxies, which show rigid-body growth in the RCs. 
Fig. 1 includes some barred galaxies, which also show similar 
rotation properties to normal galaxies. Recent CCD observations of 
the H$\alpha$-line have also revealed the nuclear rises of rotation 
velocity for many galaxies (Rubin et al. 1997; Sofue et al. 1997, 
1998, 1999; Bertola et al. 1998), in agreement with the CO results. 
It is also interesting to note that the rotation velocities in many 
galaxies do not appear to decline to zero at the nuclei, indicating 
that the mass density increases steeply toward the center (Bertola 
et al. 1998; Sofue et al. 1999).

--Fig. 1--

The fact that almost all the galaxies show steep nuclear rises
would imply that the high velocities are not due to the non-circular 
motions in bars. When observing a bar, the probability of looking it 
at a side-on view is fairly higher than that of an end-on view.
If the influence of the non-circular motion is significant, the RC
will indicate lower and more gradually rising velocity than the 
circular velocity, rather close to the pattern speed of the bar. 
Therefore, we may assume that the RCs trace the circular motions in 
the first approximation. If there exist bars, we may still underestimate 
the circular velocity, which would result in an underestimated mass 
density. 

Radial profiles of the surface-mass density (SMD) and the surface 
luminosity can be used to calculate the $M/L$ directly. As for SMD 
profiles, Bosma (1981b) used two methods, each described by Nordsieck 
(1973) and Shu et al. (1971), to derive them directly from RCs. He 
compared these methods and found the good agreement between the two 
SMD profiles. However, most 
of the current investigations have been done on the assumption that 
the visible part of a galaxy consists of a given set of a bulge and a 
disk, each assumed to have a constant $M/L$. Although this idea is 
useful to quantify the averaged characteristics in each component, it 
would be necessary to confirm whether the assumption of the constant 
$M/L$ is reliable. Kent (1986) has extensively used the ``maximum-disk 
method'' to fit RCs calculated on this assumption to observed RCs:
He has derived averaged $M/L$s in the individual components.
However, this method is based on the assumption that there is no dark 
matter in the bulge and disk. Moreover, results of such a fitting 
process much depend on bumps and wiggles in the luminosity profiles 
(Bosma 1998) as well as on the fitting range.

Forbes (1992) have derived the radial variation of the ratio of the 
total mass to total luminosity involved within a radius, $r$, which
can be called an 'integrated $M/L$'. In this paper, we will derive 
both the SMD and the surface luminosity as differential quantities 
at a given radius, and compare them directly in order to discuss the 
detailed radial distribution of the $M/L$.

In section 2, we derive the SMD profiles. We obtain the luminosity 
profiles in section 3,  describe the obtained $M/L$ profiles in section 
4, and discuss the results in section 5.

\section{Radial Profiles of the Surface-Mass Density}

We assume that the 'true' mass distribution in a real disk galaxy will 
be between two extreme cases; spherical and axisymmetric flat-disk 
distributions. We can calculate the SMD, $\sigma(R)$, directly from a 
RC for these extreme cases. 

\subsection{Spherical Mass Distribution}

The mass $M(r)$ inside the radius $r$ is given by
\begin{equation}
M(r)=\frac{r {V(r)}^{2}}{G},
\end{equation}
where $V(r)$ is the rotation velocity at $r$. Then the SMD
${\sigma}_{\rm S}(R)$ at $R$ is calculated by,
\begin{eqnarray}
{\sigma}_{\rm S}(R) & = & 2 \int\limits_0^{\infty} \rho (r) dz , \\
 & = & \frac{1}{2 \pi} \int\limits_R^{\infty} 
\frac{1}{r \sqrt{r^2-R^2}} \frac{dM(r)}{dr}dr .
\end{eqnarray}
Here, $R$, $r$ and $z$ are related by $r=\sqrt{R^2+z^2}$, and
the volume mass density $\rho(r)$ is given by
\begin{equation}
\rho(r) =\frac{1}{4 \pi r^2} \frac{dM(r)}{dr}.
\end{equation}
 
For a given RC, $V(r)$, eq. (3) can be calculated numerically.
Note, nevertheless, an underestimation of eq. (3) becomes significant
when $R \sim R_{\rm max}$, the outermost measured radius of the RC, 
because eq.(3) can be integrated only up to the $R_{\rm max}$, instead 
of $\infty$. This edge effect is negligible elsewhere because of the 
factor ${(\sqrt{r^2-R^2})}^{-1}$, by which the contribution at $r$ far 
from $R$ decreased rapidly. In Fig. 2 we show the several results 
of the integration in eq. (3) for the Miyamoto-Nagai potential 
(Miyamoto \& Nagai 1975),  by taking several different values of 
$R_{\rm max}$. Here, the `true' value for this potential is given by  
${\sigma}_{\rm D}(R)$. The figure demonstrates that the edge effect is 
almost negligible except at radii $r>0.9 R_{\rm max}$, where the SMD 
is underestimated.

--Fig. 2--

\subsection{Flat-Disk Mass Distribution}

The SMD for a thin flat-disk, ${\sigma}_{\rm D}(R)$, is derived by 
solving the Poisson's equation:
\begin{equation}
{\sigma}_{\rm D}(R) =\frac{1}{{\pi}^2 G} \left[ \frac{1}{R} \int\limits_0^R 
{\left(\frac{dV^2}{dr} \right)}_x K \left(\frac{x}{R}\right)dx + 
\int\limits_R^{\infty} {\left(\frac{dV^2}{dr} \right)}_x K \left
(\frac{R}{x}\right) \frac{dx}{x} \right],
\end{equation}
where $K$ is the complete elliptic integral and becomes very large
when $x\simeq R$ (Binney \& Tremaine 1987). 

When we calculate the ${\sigma}_{\rm D}(R)$, the following three points
must be taken into account. ({\it a}) First, eq. (5) is subject to 
the boundary 
condition, $V(0)=V(\infty)=0$. The assumption $V(0)=0$ may not be 
applied, if there is a black hole (BH) at the galactic center. However, 
a central BH with mass of several to hundreds million $M_{\odot}$ 
dominates the RC only within a few pc, and it does not influence the 
SMD except at the center. On the other hand, the spatial resolution 
in our data is of the order of 100 pc, thus the RCs could hardly trace 
the central BH. So, for our purposes we can safely ignore its existence. 
({\it b}) Second, when calculating the first term on the right hand side 
of eq. (5), we have only a few data points in the central region within 
a radius comparable to the spatial resolution, where the reliability of 
the calculated $V(r)$ is lower than the outer region. ({\it c}) Third,  
the upper limit of the integration of the second term is $R_{\rm max}$ 
instead of infinity. Since the RCs are nearly flat or declining outward 
from $R = R_{\rm max}$, the second term becomes negative. Hence, we 
overestimate ${\sigma}_{\rm D}(R)$ at $R\simeq R_{\rm max}$, and, as 
calculated here, the SMD near the outer edges will give an upper limit 
to the true value.  For such galaxies as NGC 1068 and NGC 4569, whose 
rotation velocities appear to increase even outwards of $R = R_{\rm max}$,
the error would be still larger. 

\subsection{Verification using the Miyamoto-Nagai Potential}

We, here, examine how the results for the spherical and flat-disk 
assumptions differ from each other as well as from the `true' SMD by 
adopting the Miyamoto-Nagai (MN) potential (Miyamoto \& Nagai 1975),
which realistically comprises four components denoting a compact nucleus,
a bulge, a disk, and a dark halo. 
Given a set of parameters for the MN potential suitable for our Galaxy,
we can calculate the `true' SMD as well as the RC. Using this RC, we 
derive the SMD both for a spherical and flat-disk assumptions using 
the methods as in the previous sections. The MN potential is given by  
$\Phi = \sum_{i=1}^{4} G M_{i} {[r^{2} + {\{a_{i} + {(z^{2}
+ b_{i}^{2})}^{0.5}\}}^{2}]}^{-0.5}$
and $(a_{1-4},b_{1-4},M_{1-4}) = (0.00,0.12,0.05)$,
$(0.00,0.75,0.10)$, $(6.00,0.50,1.60)$, $(15.00,15.00,3.00)$ in the 
units of $({\rm kpc},{\rm kpc},10^{11}M_{\odot})$.

In Fig. 3 we show the calculated true SMD, and the SMDs for a spherical
and a flat-disk assumptions calculated from the RC. We used the RC 
only up to $R= 20$ kpc. Fig. 3 shows that the spherical case well 
reproduces the true SMD for the inner region. This is not surprising, 
because a spherical component is dominant within the bulge. On the 
other hand, the flat-disk case mimics the true one in the disk region, 
which is also reasonable. Near the outer edge, the flat-disk case also 
appears to be a better tracer of the true SMD. We stress that the 
results from the two extreme assumptions, spherical and flat-disk, 
differ at most by a factor of two, and do not differ by more than a 
factor of 1.5 from the true SMD. We may thus safely assume that the 
true SMD is in between the two extreme cases: It is better represented 
by a spherical case for the inner region, while  a flat-disk case is 
better for the disk and outer part.

--Fig. 3--

\subsection{The Data and Results}

We use the RCs in Sofue (1997) to derive the SMD profiles. We select 
18 SA and SAB spiral galaxies, for which the spatial resolution (FWHM) 
is less than 650 pc. Fig. 4 shows the calculated ${\sigma}_{\rm S}(R)$ and 
${\sigma}_{\rm D}(R)$ with error-bars, respectively. Among the sample, NGC 
4527 has the highest signal-to-noise ratio, which have been obtained 
from CO-, HI- and \ha-line data. The parameters of the sample are listed 
in Table 1.

--Table1--

--Fig. 4--

From Fig. 4 the typical difference of SMD values between the two 
assumptions (sphere and flat-disk) is estimated to be $\pm$0.05 to 
0.1 in a logarithmic scale. The error due to the error in rotation 
velocities is about $\pm$0.05. Hence, the effective error of SMD 
calculation is estimated to be $\pm$0.1 to 0.15. 

In Figs. 5 and 6 the SMD profiles of Sb and Sc galaxies are shown 
together. Although some Sc galaxies distinctly have smaller SMD values 
than the average, the difference in SMD between the two galaxy types 
is fairly small. Namely, Sb and Sc galaxies show similar surface-mass 
distributions. In the disk region at radii 3 to 10 kpc the SMD decreases 
exponentially outward by about 0.3 times/5kpc. In the bulge region 
($<$ 2 kpc) of some galaxies, the SMD shows a power-law decrease, much 
steeper than an exponential decrease. This may indicate that the 
surface-mass concentration in the central region is higher than 
the luminosity concentration. The central activities appear to be not 
directly correlated with the mass distribution. Note that the present 
sample includes galaxies showing Seyfert (NGC  1068, NGC  4569), LINER 
(NGC  3521, NGC  4569), jets (NGC 3079), and black holes (the Milky Way).

--Fig. 5a,b--

--Fig. 6a,b--

\section{Luminosity Profiles}

\subsection{Data Selection}

Luminosity profiles in bulge regions are affected by the interstellar
extinction. The extinction is higher nearer the center; it is higher 
for a larger inclination angle. It is also higher in shorter wave 
ranges. In fact, the color-gradient ($B-K$ or $V-K$) between the bulge 
and disk for galaxies with larger inclinations ($i \geq 60^{\circ}$)
is greater than that for lower-inclination galaxies (H\'{e}raudeau 
et al. 1996). On the other hand, galaxies with $i \leq 51^{\circ}$ 
in de Jong \& van der Kruit (1994) do not show significant differences 
in the color gradient. 

In the present analysis, we have selected the galaxies whose 
inclinations are less than ${70}^{\circ}$ if they are observed in 
NIR-bands and $i$-band, and the galaxies with inclinations less than 
${60}^{\circ}$ if their data are in optical. In the latter case their 
mean surface luminosities may be weakened by 1 mag ${\rm arcsec}^{-2}$ at 
most, depending on the  inclination. We have, thus, selected 11 galaxies 
out of the 19 RC-selected galaxies.

\subsection{Description of the Data}

Table 1 lists the parameters for individual galaxies and their 
references. Since the data have been reduced in different ways by 
different authors, we have checked if their data are suitable for 
the present analysis of $M/L$. We briefly describe the data, particularly 
how the surface-luminosity profiles, $\mu(r)$, have been obtained.

(a) Regan \& Vogel (1994) have de-projected NGC 598, and calculated 
$\mu(r)$  mag ${\rm arcsec}^{-2}$ by averaging azimuthally in rings of 
width ${\pm 3}^{''}$. The deviation from axi-symmetry is at most about 
0.1 mag ${\rm arcsec}^{-2}$ including the observational error, which is much 
smaller than the error of the SMD. The effects by spiral arms are small. 

(b) Kodaira et al. (1990) and (d) de Jong \& van der Kruit (1994) 
measured $\mu(r)$ along the major axes. The errors of $\mu(r)$ are as 
small as $\pm 0.1-0.15$ mag ${\rm arcsec}^{-2}$. 

(c) Frei et al. (1996) published imaging data of spiral galaxies in FITS 
format. We have used them to derive $\mu(r)$ at the same position angles 
at which the RCs have been measured. The errors of $\mu(r)$ are found to 
be smaller than that of SMD. 

(e) Boselli et al. (1997)  reduced their data in the same way as Regan 
\& Vogel (1994) to derive luminosity profiles. For NGC  4569 they assumed 
an inclination of 67${}^{\circ}$ and a position angle of 23${}^{\circ}$, 
which are almost the same as those we adopt here. The errors of $\mu(r)$ 
are less than 0.1 mag ${\rm arcsec}^{-2}$ between $r = 0{''}$ and $150{''}$. 

(f) M\"{o}llenhoff et al. (1995) calculated $\mu(r)$ of NGC  4736 along 
the semi-major axis for different position angles and ellipticities. 
Because of the central weak bar, $\mu(r)$ may be about 0.5 
mag ${\rm arcsec}^{-2}$ brighter at $r = {(10 \pm 5)}^{''}$ than that at 
position angle of ${108}^{\circ}$ as adopted for the RC. 

We may, thus, consider that the errors in the luminosity profiles caused 
by the different ways is negligible compared to that due to the 
observations, except for NGC 4736. We note, however, that observations 
in the V-band (NGC 4303, NGC  5055, NGC  6946) may be affected by spiral 
arms, and the errors will amount to about $\pm$0.5 mag ${\rm arcsec}^{-2}$ 
($\sim\pm$0.2 in the unit of log($L_{\odot}$ ${\rm pc}^{-2})$).

\section{Mass-to-Luminosity Ratios}

To obtain $M/L$ profiles, we first converted the unit of $\mu(r)$ 
from mag ${\rm arcsec}^{-2}$ into $L_{\odot}$ ${\rm pc}^{-2}$, then divided 
the SMD by $\{\mu(r)$($L_{\odot}$ ${\rm pc}^{-2}$)$\times$cos$i\}$ 
corrected for the inclination. We adopt the following solar luminosities 
in the individual bands: $L_{\odot ,V}=4.83$ mag, $L_{\odot ,K}=3.41$, 
$L_{\odot ,K^{'}}=3.45$ and $L_{\odot ,i}=4.83$. We note, however, that 
we cannot compare the absolute values of $M/L$ profiles directly among 
the galaxies, because we use different filters for different galaxies.
Hence, we here focus on relative variations of $M/L$ with radius in each 
individual galaxy rather than compare the absolute $M/L$ values among 
different objects. Fig. 4 shows the thus 
obtained $M/L$ profiles for the 11 galaxies, and Fig. 7 shows the same 
but for the inner 4 kpc region. As the errors of SMD are much larger 
than those of $\mu(r)$ for all the galaxies, only the errors in SMD 
are indicated.

--Fig. 7--

\subsection{Results for Individual Galaxies}

We discuss the results for the individual galaxies:

{\bf NGC 598:} 
This is a small-mass galaxy which has a mildly rising RC to reach only 
about 100  km s$^{-1}$  in the disk region. The $M/L$ increases slowly 
by 1.7 times from $r =$ 1 to 3 kpc. In the bulge region it appears 
to decrease inward rapidly, with large error-bars. It also appears to 
increase inward within 150 pc but it is not conclusive, because the 
spatial resolution is 210 pc.

\noi{\bf NGC 1068:} 
The $M/L$ in the disk region increases slowly at 4.6 times from $r =$ 2 
to 6 kpc. In the bulge region it decreases more rapidly (0.28 times 
from $r =$ 2 kpc to 800 pc) than in the disk, but within $r =$ 800 pc 
turns to be nearly flat or somewhat increasing. This is a Seyfert 
type-II galaxy, so the luminosity profile in V-band may be not exact 
in the central region.

\noi{\bf NGC 2903:} 
Both the SMD profile and $\mu(r)$ have a remarkable bump at $r\sim$ 2 kpc. 
Since the shape of the RC is also peculiar, we may not neglect the effect 
of non-axisymmetric component such as due to a spiral arm. Interestingly, 
however, there appears no corresponding bump in the $M/L$. The $M/L$ in 
the disk region increases slowly by 5.5 times from $r =$ 4 to 9 kpc. 
This Sc galaxy has a compact bulge and we cannot evaluate the $M/L$ of 
the inner bulge for the insufficient spatial resolution.

\noi{\bf NGC 4303:} 
Because of some irregularities caused by spiral arms, the $M/L$ in the 
disk region behaves irregularly, although it increases outward on the 
average. In the compact bulge region the $M/L$ decreases inward steeply 
from $r =$ 2 kpc to 500 pc. It, then, turns to increase within $r =$ 200 
pc toward the center.

\noi{\bf NGC 4321:} 
The $M/L$ in the disk increases slowly by 2.0 times from $r =$ 3 to 8 kpc. 
In the bulge region it seems to decrease inward and then increase near
the center, though the error-bar is large. Some luminous bumps of about 
0.5 mag ${\rm arcsec}^{-2}$ are found at $r =$ 700 pc, 5 kpc and at 11 kpc, 
which may be caused by spiral arms. The $M/L$ increases systematically 
inward by about 2 times from $r =$ 300 to 150 pc.

\noi{\bf NGC 4527:} 
The $M/L$ in the disk region is almost flat, or slowly increases outward. 
In the bulge region it decreases inward from $r =$ 2 kpc to 500 pc. Then, 
it turns to increase steeply toward the center.

\noi{\bf NGC 4569:} 
The $M/L$ in the disk region increases by 2.4 times from $r =$ 2 to 
5 kpc. In the bulge region it appears to decrease inward,  and then turns 
to increase toward the center. However, the spatial resolution is not 
sufficient to confirm the central increase. 

\noi{\bf NGC 4736:} 
The luminosity profile is obtained only in the bulge region. The $M/L$ 
in the bulge region decreases distinctly inward by 0.38 times from $r =$ 
1.8 kpc to 800 pc. The $M/L$ in the inner-most region cannot be evaluated 
because of the large spatial resolution.

\noi{\bf NGC 5055:} 
The $M/L$ in the disk region increases slowly by 3.0 times from $r =$ 5 
to 15 kpc. In the bulge region its variation is difficult to evaluate 
because of the insufficient resolution as well as due to the large error 
in the luminosity profile by spiral arms. It looks flat in the outer 
bulge and increases inward in the inner bulge region.

\noi{\bf NGC 5194:} 
The $M/L$ in the disk region is almost flat or rather decreases outward. 
The peculiar behavior of the SMD, and therefore of the $M/L$, is probably 
due to a disturbed RC by the tidal interaction with the companion galaxy 
NGC 5195. In the bulge region, the $M/L$ is flat or slightly decreases 
inward to $r =$ 700 pc. Within $r =$ 700 pc, it looks flat or slightly 
increasing, while the spatial resolution is not sufficient to reveal the 
details.

\noi{\bf NGC 6946:} 
The SMD profile for this galaxy has been calculated with the highest 
accuracy among the 11 galaxies of the present sample (FWHM of 110 pc).
The $M/L$ in the disk region is flat from $r =$ 1 to 3 kpc and, 
then, it increases slowly by 2.8 times from $r =$ 3 to 10 kpc. In 
the compact bulge region, it shows a remarkable increase toward the 
center by about an order of magnitude from $r =$ 1 kpc to 50 pc.

\subsection{General Characteristics}

From Fig. 4 and Fig. 7 we summarize the general characteristics of the
$M/L$ behavior as follows: 

\noindent (1) The $M/L$ remains nearly constant in the inner disk. 

\noindent (2) It increases gradually outward in the outer disk for all the 
galaxies, except for NGC 5194, which is tidally disturbed by the companion. 

\noindent (3) The $M/L$ increases more steeply in the outer halo, 
indicating the massive dark halo. 

\noindent (4) Bulge components show a constant or a slightly smaller value 
than the disk value. 

\noindent (5) For NGC 4527 and NGC 6946 the $M/L$ increases steeply toward 
the nucleus inside the bulge. 

This drastic increase of $M/L$ within $r =$ 500 pc for NGC 4527 and 
NGC 6946 is of particular interest: For NGC 4527 this increase 
amounts to $4.9\pm1.0$ times/350pc from $r =$ 500 to 150 pc, and
for NGC 6946, $3.5\pm0.9$ times/500pc from $r =$ 600 to 100 pc. 
Since the results are based on optical data, these values may be 
overestimated because of optical extinction. However, it is known 
that color gradients like $V-K$ and $r-K$ of face-on galaxies are of 
the order of 1 mag ${\rm arcsec}^{-2}$ (e.g., de Jong \& van der Kruit 1994; 
Terndrup et al. 1994). Therefore, we may conclude that the increase of 
$M/L$ for NGC 6946, whose inclination is only 30${}^{\circ}$, is 
significantly greater than that expected from the optical extinction 
effect. For NGC 4527 with $i=60^{\circ}-75^{\circ}$, the $B-{K}^{'}$ 
color gradient is estimated to be about 1.5 mag ${\rm arcsec}^{-2}$ 
(H\'{e}raudeau et al. 1996). In the present data, the luminosity profile 
of NGC 4527 is in the Thuan-Gunn $i$-band, and we may conclude that 
the central increase of $M/L$ for NGC 4527 will be also significant. 

Such a remarkable increase of $M/L$ toward the center is, however, 
not detected in the other galaxies. For NGC 1068, NGC 2903, NGC 4303, 
NGC 4569, NGC 4736 and NGC 5194, the spatial resolution is not 
sufficient to reveal such an increase even if it did exist. For NGC 4321 
and NGC 5055 we see some hints of a possible central increase. For the
 small-mass galaxy, NGC 598, we cannot find any increase inward near 
the center. 

In current studies of the $M/L$ of spiral galaxies, the bulge and disk 
$M/L$s have been assumed to be radially constant. However, it does not 
necessarily mean that the total $M/L$ value is constant in the region 
corresponding to each component: In Fig. 8 we present the variations 
of the $M/L$ for galaxies with inclinations less than $80^{\circ}$.
Here, the $M/L$ value is normalized to unity at ${\it r}=2$ kpc. 
It is now obvious that the total $M/L$ is not constant at all 
within a galaxy, but it varies significantly in the bulge and disk, 
not only in the massive halo. The $M/L$ value often increases outward 
by 100 times/(10$-$30) kpc from the disk to the outer halo.
Only one exception is NGC 5194, whose peculiar behavior of the $M/L$ 
would be due to the tidal interaction with the companion. 

--Fig. 8--

\section{Discussion}

\subsection{Massive Dark Core}

We have used high-accuracy RC data to derive SMD profiles for 19 
galaxies and $M/L$ profiles for 11 galaxies. For three galaxies, NGC 
598, NGC 4527, and NGC 6946, we obtained both high spatial-resolution 
($r \leq 200$ pc) RCs and luminosity profiles without large 
irregularities due to spiral arms and/or bars. Among these three 
galaxies, two galaxies, NGC 4527 and NGC 6946, are found to show a 
steep inward increase of $M/L$ in the central region, $r \leq$ 500 pc. 
Such a steep central increase of $M/L$ may imply that the bulges contain 
an excess of (dark) mass inside a few hundred pc, which we call a 
``massive dark core'' with a scale radius of 100 $-$ 200 pc. This 
yields a mass of $M \sim RV^{2}/G \sim 10^{9}M_{\odot}$, assuming 
$V(R=100-200 {\rm pc})=200$ km s$^{-1}$. The massive dark 
core could be an object linking the galactic bulge with a massive black 
hole in the nuclei (Miyoshi et al. 1995; Genzel et al. 1997; Ghez et al. 
1998) and/or massive core objects causing a central Keplerian RC 
(Bertola et al. 1998). 

For the Sbc galaxy NGC 4527 we have detected both a dark core and a
normal bulge within the luminous bulge region.
For the Scd galaxy NGC 6946 we found the dark core, 
whereas we could hardly detect a normal bulge component.
On the other hand, we could see only a normal bulge component in the 
Scd galaxy NGC 598.
Such a difference may indicate that the existence of a dark
core may not be directly related to the morphology. 
It is also interesting to note that 
the scale sizes of the dark cores in the two galaxies may not be uniquely
correlated with the size of luminous bulges.
Although dark cores will have a crucial implication for the formation, 
evolution and dynamics of the central bulges of galaxies,
their origin and universal properties remain open to discussion. 

\subsection{Dark-Mass Fraction}

It is interesting to derive the distribution of dark mass fraction 
(hereafter DMF), $F_{\rm DM}$, defined by the local dark-to-total 
mass density ratio at a given $r$.  
If the luminosity profile can be corrected for the extinction, we can
define the following quantity:
\begin{equation}
\frac{{\sigma}_{\rm total}}{{\mu}_{*}}(r)=\frac{{\sigma}_{*}+{\sigma}_{\rm dark}}{{\mu}_{*}},
\end{equation}
where
${\sigma}_{\rm total}$, ${\sigma}_{*}$, ${\sigma}_{\rm dark}$, and ${\mu}_{*}$ 
denote the SMD of the total mass, visible mass, dark mass, 
and the surface-luminosity density, respectively. 
If the $M/L$ of stars on the average at a given {\it r}, which will be
derived from luminosity profiles by multi-band observations in optical
and infrared bands, is known, then we can define the DMF by
\begin{equation}
{F_{\rm DM}}(r)=\frac{{\sigma}_{\rm dark}}{{\sigma}_{\rm total}}=\frac{{\sigma}_{\rm total}/{\mu}_{*}(r)-{\sigma}_{*}/{\mu}_{*}(r)}{{\sigma}_{\rm total}/{\mu}_{*}(r)}.
\end{equation}
This quantity can be used to evaluate the dark-matter distribution. 

Since the absolute calibration of the stellar $M/L$ and the precise
correction for the extinction are beyond our scope, here we only briefly 
attempt to examine qualitative trends of variation of the DMF.
We examine the following two cases: In one case, we assume 
${\sigma}_{*}/{\mu}_{*}(r)$ to be constant and equal to 
${\sigma}_{\rm total}/{\mu}_{*}(r)$ in the inner disk region;
in the other case, we take it equal to half of that.
We thus derive $F_{\rm DM}$({\it r}) and ${\sigma}_{\rm dark}$({\it r})
for NGC 6946: the results are shown in Figs. 9a and 9b,
corresponding to the above two cases, respectively. 
Since ${\sigma}_{*}/{\mu}_{*}(r)$ will be smaller in the outer disk region 
owing to the dominance of younger stars of lower metallicity 
(e.g., de Jong 1996), we may safely consider that the DMF is
at least 80 $-$ 90 \% in the outer disk beyond 8 kpc.
In the inner disk region ($r$ $\sim$ 4$-$5 kpc) the DMF appears to 
be more than 50 \%. However, the uncertainty is too large 
to give a more detailed discussion. 

--Fig. 9a,b--

\subsection{Dark Mass v.s. Visible Mass in the Disk Region}

The radial distribution of DMF can be used to derive individual 
distributions of dark mass and visible mass in the disk region. We 
attempt the following analysis: 

(a) We fit the observed luminosity profiles in a logarithmic scale by a 
non-linear least square fitting by a model, which consists of an 
exponential bulge and an exponential disk. The model is given by
\begin{equation}
{\mu}_{\rm model}(r) = {\mu}_{\rm disk}(r) + {\mu}_{\rm bulge}(r),
\end{equation}
where
\begin{equation}
{\mu}_{\rm disk}(r) = {\mu}_{\rm e,disk}{\rm e}^{-1.679(r/{r}_{\rm e,disk}-1)}
\end{equation}
and
\begin{equation}
{\mu}_{\rm bulge}(r) = {\mu}_{\rm e,bulge}{\rm e}^{-1.679(r/{r}_{\rm e,bulge}-1)}.
\end{equation}
We could successfully fit the data by this model for nine galaxies, except 
for NGC 4736 and NGC 5194: NGC 4736 was short of the data in the disk region, 
and NGC 5194 does not have a disk described by the exponential function 
probably because of the tidal interaction. 

(b) We define boundary radius, $r_{0}$, at which 
${\mu}_{\rm disk}/{\mu}_{\rm bulge}=10$, and assume that the $M/L$ of 
visible mass, ${\sigma}_{*}/{\mu}_{*}$, is constant outside of $r_{0}$. 
We will focus on the disk region in the following. 

(c) We set the $M/L$ value at $r = r_{0}$ to that of the total mass, 
so that the visible mass becomes maximum in each galaxy.
From the consideration in section 2.3., we adopt a flat-disk mass 
distribution for SMDs. By subtracting the SMDs of visible mass from those
of total mass, we obtain distributions of dark mass for the nine galaxies.
We define, ``mass-boundary radius'',  $r_{\rm mb}$, outside of which the 
fitted SMDs of visible mass is dominated by the dark mass.
Note that, because of the assumption of maximum $M/L$ of visible mass,
we, in fact, derive the maximum mass-boundary radius, $r_{\rm mb,max}$, 
inside of which dark mass may even exceed visible mass.

In Fig. 10, we show the results of SMD profiles of the total, visible, 
and dark mass for each galaxy. The parameters are listed in Table 2. Two 
interesting points are found: First, dark mass tends to dominate visible 
mass even in the inner disk region. In Fig. 11, the relation between the 
effective radius of a disk component, ${r}_{\rm e,disk}$, and $r_{\rm mb,max}$ 
is shown. As $r_{\rm mb,max}$ is an upper limit of $r_{\rm mb}$, 
${r}_{\rm e,disk}$ is approximately equal to or less than 
$r_{\rm mb,max}$. This trend appears to be independent of morphology, 
partly because the sample is biased to the late-type galaxy.
Second, there may be a linear correlation between the two parameters 
in Fig. 11. This trend is caused by the maximum $M/L$ of visible mass
in a disk. But the $M/L$s adopted here range
from 0.5 to 4.0 except for NGC 6946, and even if they are set to a half,
$r_{\rm mb,max}$ will be reduced only by a few kpc. This correlation 
may indicate that, if any, dark mass traces visible 
mass only up to $r = {r}_{\rm e,disk}$ at most, and that the luminous scale of
galaxies is almost independent of the distribution of a dark halo, because 
the difference in SMD profiles is small (see Fig. 5 and 6).

--Fig. 10--

--Fig. 11--

--Table 2--

\acknowledgments

\newpage
\parindent=0pt
{\bf Figure Captions}

{\bf Fig. 1:} Top: The rotation curves of 15 Sb (solid lines) and 3 
SBc (dashed) galaxies. 
Bottom: Those of 15 Sc (solid lines) and 2 SBc (dashed) galaxies. 
Note that small-mass galaxies, whose maximum disk velocities are 
100 to 200 km s$^{-1}$, tend to show mild rises. 

{\bf Fig. 2:} Surface-mass-density profiles calculated from a 
rotation curve given by the Miyamoto-Nagai potential of our Galaxy  
under an assumption of spherical mass distribution. 
The solid, dashed, and dot-dashed line-profiles represent results for 
different maximum radii, $R_{\rm max}$ = 20, 18 and 16 ${\rm kpc}$, 
respectively, for the upper limit of the integral in eq. (3) 
in place of $\infty$.

{\bf Fig. 3:} Top: The real (solid line) SMD, and spherical (dashed),
and flat-disk (dot-dash) SMD profiles calculated for the rotation
curve of the Miyamoto-Nagai potential.
The upper limit of the integration is taken as
$R_{\rm max} = 20$ kpc in place of $\infty$.
Bottom: The rotation curve by the Miyamoto-Nagai model of our Galaxy.

{\bf Fig. 4:} Top: (S) Surface-mass-density profile for the spherical 
assumption.
(D) Same but for the flat-disk assumption. 
An bold line denotes the luminosity profile, corrected for the inclination.
Middle: Mass-to-Luminosity ratio for (S) spherical and (D) flat-disk 
assumptions. 
Bottom: Rotation curve.
The error-bars are caused mainly by those of rotation curve.
The true values of the surface-mass-density  will lie in the 
shaded area between (S) and (D). The horizontal error-bar in the top-left 
corner denotes the angular (spatial) resolution of the surface-mass-density
 (FWHM), which is larger than that of the luminosity profile in all the cases. 
The interval of the shading lines corresponds to a double of
the interval of data sampling points.

{\bf Fig. 5a:} Top: Surface-mass-density profiles of Sb galaxies
calculated for the spherical assumption. 
Because of the edge effect of the integration of eq. (3), the SMD 
curves decline rapidly in the outer 2 to 3 kpc region, where 
we must not rely on the obtained SMD profiles. 
Bottom: Same but for the inner region at  $r=$ 0 to 2 kpc.

{\bf Fig. 5b:} Same as Fig. 5a, but for the flat-disk assumption. 
 
{\bf Fig. 6a:} 
Same as Fig. 5a, but for spherical assumption and Sc galaxies.
Note that some Sc galaxies show smaller SMD values than the total 
average of our sample.
The Scd galaxy NGC 598 shows the smallest values in $r$ = 0 $\sim$ 
6 kpc. The peculiar behavior of NGC 5194 at $r\sim$ 8 kpc may be
due to the tidal interaction with the companion. 

{\bf Fig. 6b:} 
Same as Fig. 6a, but for the flat-disk assumption. 

{\bf Fig. 7:} $M/L$ profiles at $r=$ 0 to 4 kpc. (a): Calculated for 
the spherical assumption. (b): For the flat-disk assumption. 
The meanings of the horizontal and vertical error-bars 
and the shaded area are same as those in Fig. 4.

{\bf Fig. 8:} $M/L$ distributions normalized at ${\it r}=2$ kpc, as
obtained by using the SMD and the V-band photometric data of 
Kodaira et al. (1990).

{\bf Fig. 9a:} Top: The distribution of the dark mass fraction (DMF)
 in NGC 6946 for a case of ${\sigma}_{*}/{\mu}_{*}(r)=10$. The error-bars are 
due to the difference between the assumptions of spherical and flat-disk mass 
distributions. Note that the DMF is more than 50 \% even in the disk region. 
Bottom: The SMD distribution of the dark mass (solid line)
and the total mass (dashed). 

{\bf Fig. 9b:} The same as Fig. 9a, but for the 
${\sigma}_{*}/{\mu}_{*}(r)=5$. Note that more than half of the total mass
is dominated by dark mass in $r \geq 5$ kpc not strongly depending on the 
${\sigma}_{*}/{\mu}_{*}(r)$.

{\bf Fig. 10:} SMD profiles of the total mass (dot-dash line), 
visible mass (solid) and dark mass (dashed). The narrow lines are 
based on the two component model, while bold lines are directly 
calculated from the data.

{\bf Fig. 11:} The relation between ${r}_{\rm e,disk}$ and $r_{\rm mb,max}$.
The straight line shows ${r}_{\rm mb,max} = r_{\rm e,disk}$.
 
\newpage

\begin{table}[h]
\begin{center}
\begin{tabular}{llccccccl} \hline\hline
\multicolumn{1}{l}{(1)}&\multicolumn{1}{l}{(2)}&\multicolumn{1}{c}{(3)}&
\multicolumn{1}{c}{(4)}&\multicolumn{1}{c}{(5)}&\multicolumn
{1}{c}{(6)}&\multicolumn{1}{c}{(7)}&\multicolumn{1}{c}{(8)}&\multicolumn
{1}{l}{(9)}\\ \hline
NGC 253        & Sc    & 78.5     & 2.5  & 15             & 0.18             &         &     &   \\
IC  342        & Sc    & 25       & 3.9  & 4              & 0.076            &         &     &   \\
NGC 598 (M33)  & Sc    & 54       & 0.79 & 55             & 0.21             & K       & 2   &(a)\\
NGC 891        & Sb    & 88.3     & 8.9  & 4              & 0.17             &         &     &   \\
NGC 1068 (M77) & Sb Sy & 46       & 18.1 & 5$\times$4     & 0.44$\times$0.35 & V       & 3   &(b)\\
NGC 2903       & Sc    & 35       & 6.1  & 15             & 0.44             & $i$     & 1.2 &(c)\\
NGC 3079       & Sc    & $\sim$90 & 15.6 & 4              & 0.30             &         &     &   \\
NGC 3521       & Sbc   & 75       & 8.9  & 15             & 0.65             &         &     &   \\
NGC 3628       & Sb    & $>$86    & 6.7  & 3.9            & 0.13             &         &     &   \\
NGC 4258 (M106)& Sbc   & 67       & 6.6  & 15             & 0.48             &         &     &   \\
NGC 4303       & Sc    & 27       & 8.1  & 4              & 0.16             & V       & 3   &(b)\\
NGC 4321 (M100)& Sc    & 27       & 15   & 4              & 0.29             & K       & 1.8 &(d)\\
NGC 4527       & Sb    & 69       & 22   & 2.0            & 0.21             & $i$     & 1.2 &(c)\\
NGC 4569 (M90) & Sab   & 63       & 8.2  & 9.9$\times$4.8 & 0.39$\times$0.19 & K$^{'}$ & 2   &(e)\\
NGC 4736 (M94) & Sab   & 35       & 5.1  & 15             & 0.37             & K       & 1   &(f)\\
NGC 5055 (M63) & Sbc   & 55       & 8    & 15             & 0.58             & V       & 4   &(b)\\
NGC 5194 (M51) & Sc    & 20       & 9.6  & 13             & 0.60             & V       & 4   &(b)\\
NGC 6946       & Sc    & 30       & 5.5  & 4              & 0.11             & V       & 3   &(b)\\
Milky Way      & Sb    & 90       & 0    &                &                  &         &     &   \\ \hline
\end{tabular}
\end{center}
\caption{(1)Galaxy name, (2)morphological type, (3)inclination ($^\circ$),
 (4)distance (Mpc), (5),(6)angular (spatial) resolution of the rotation curve, 
in the units of ($''$) and (kpc), (7)photometric band, (8)angular resolution
of the luminosity profile ($''$), and (9)reference to the luminosity profile:
(a)Regan \& Vogel (1994), (b)Kodaira et al. (1990),
(c)Frei et al. (1996), (d)de Jong \& van der Kruit (1994), (e)Boselli et
 al. (1997), and (f)M\"{o}llenhoff et al. (1995). Except for NGC 4527, 
(2)$\sim$(6) are taken or calculated from Sofue (1997), while for NGC 4527 they
are from Sofue et al. (1998). (7) and (8) are taken from their references. 
(5), (6) and (8) are shown as FWHM.}
\end{table}

\newpage

\begin{table}[h]
\begin{center}
\begin{tabular}{lccccccc} \hline\hline
\multicolumn{1}{l}{Galaxy}&\multicolumn{1}{c}{${\mu}_{\rm e,disk}$}&
\multicolumn{1}{c}{$r_{\rm e,disk}$}&\multicolumn{1}{c}{${\mu}_{\rm 
e,bulge}$}&\multicolumn{1}{c}{$r_{\rm e,bulge}$}&\multicolumn{1}{c}
{$r_{0}$}&\multicolumn{1}{c}{$r_{\rm mb,max}$}&\multicolumn{1}{c}
{${(M/L)}_{\rm visible}$}\\ \hline
NGC  598 & 166. & 2.25 &  205. & 0.29 & 0.50 &  1.8 & 0.5 \\
NGC 1068 & 205. & 2.94 & 2926. & 0.63 & 2.39 &  3.5 & 2.5 \\
NGC 2903 & 164. & 2.89 &  575. & 0.35 & 0.85 &  2.0 & 3.0 \\
NGC 4303 &  44. & 2.90 & 1258. & 0.18 & 0.65 &  2.6 & 3.0 \\
NGC 4321 & 187. & 5.86 & 3265. & 0.53 & 1.79 &  7.2 & 1.5 \\
NGC 4527 &  67. & 7.62 &  552. & 0.70 & 2.03 & 10.1 & 1.5 \\
NGC 4569 & 243. & 3.41 & 6530. & 0.31 & 1.13 &  2.1 & 1.0 \\
NGC 5055 &  39. & 6.06 &  196. & 1.00 & 2.83 &  8.0 & 4.0 \\
NGC 6946 &  26. & 4.65 &   96. & 0.21 & 0.48 &  4.7 & 8.0 \\ \hline
\end{tabular}
\end{center}
\caption{Fitting parameters. Column 2 and 4 are in the unit of 
($L_{\odot}$ ${\rm pc}^{-2}$)}, the rest are in the unit of (kpc). 
\end{table}

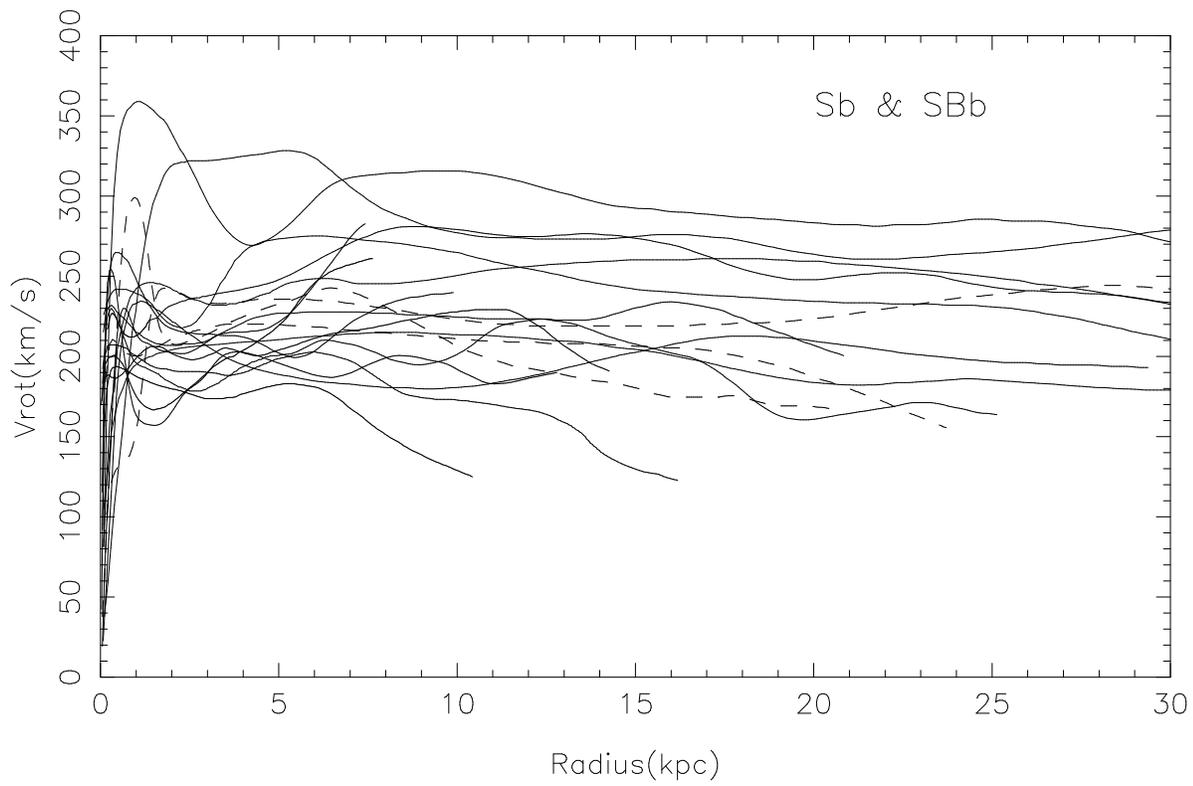
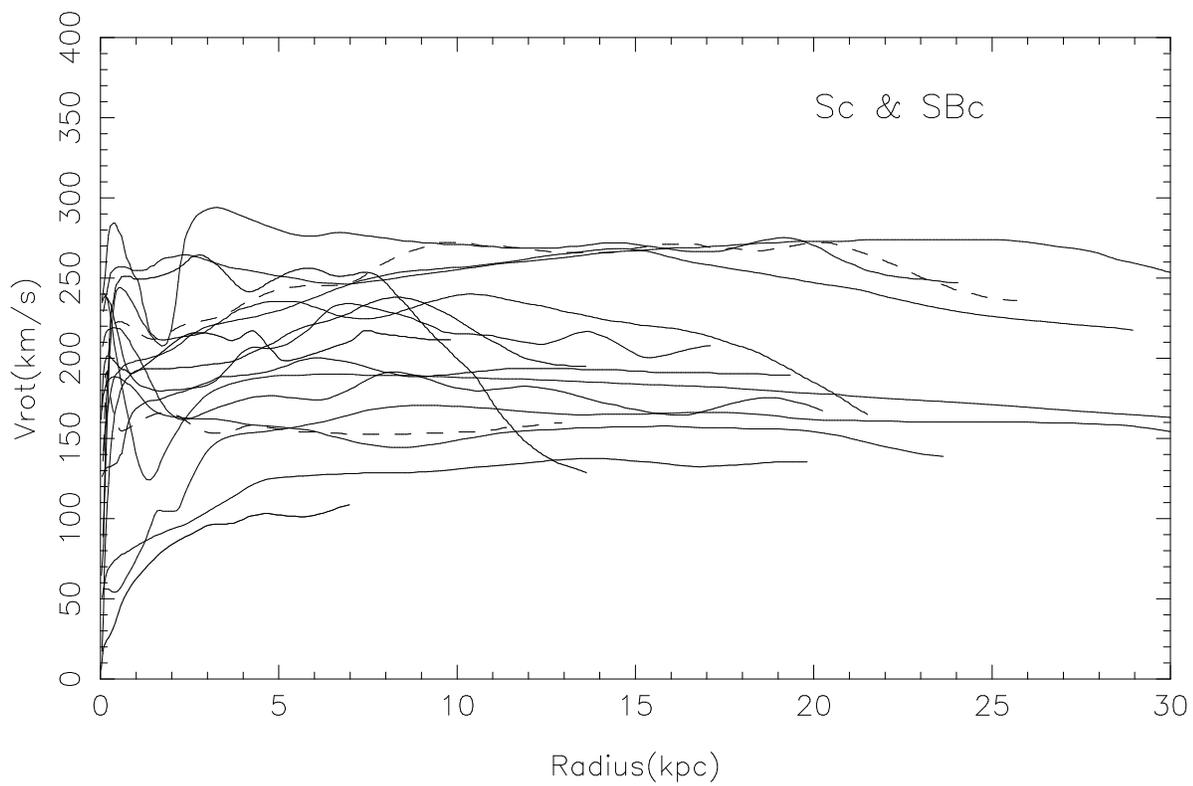
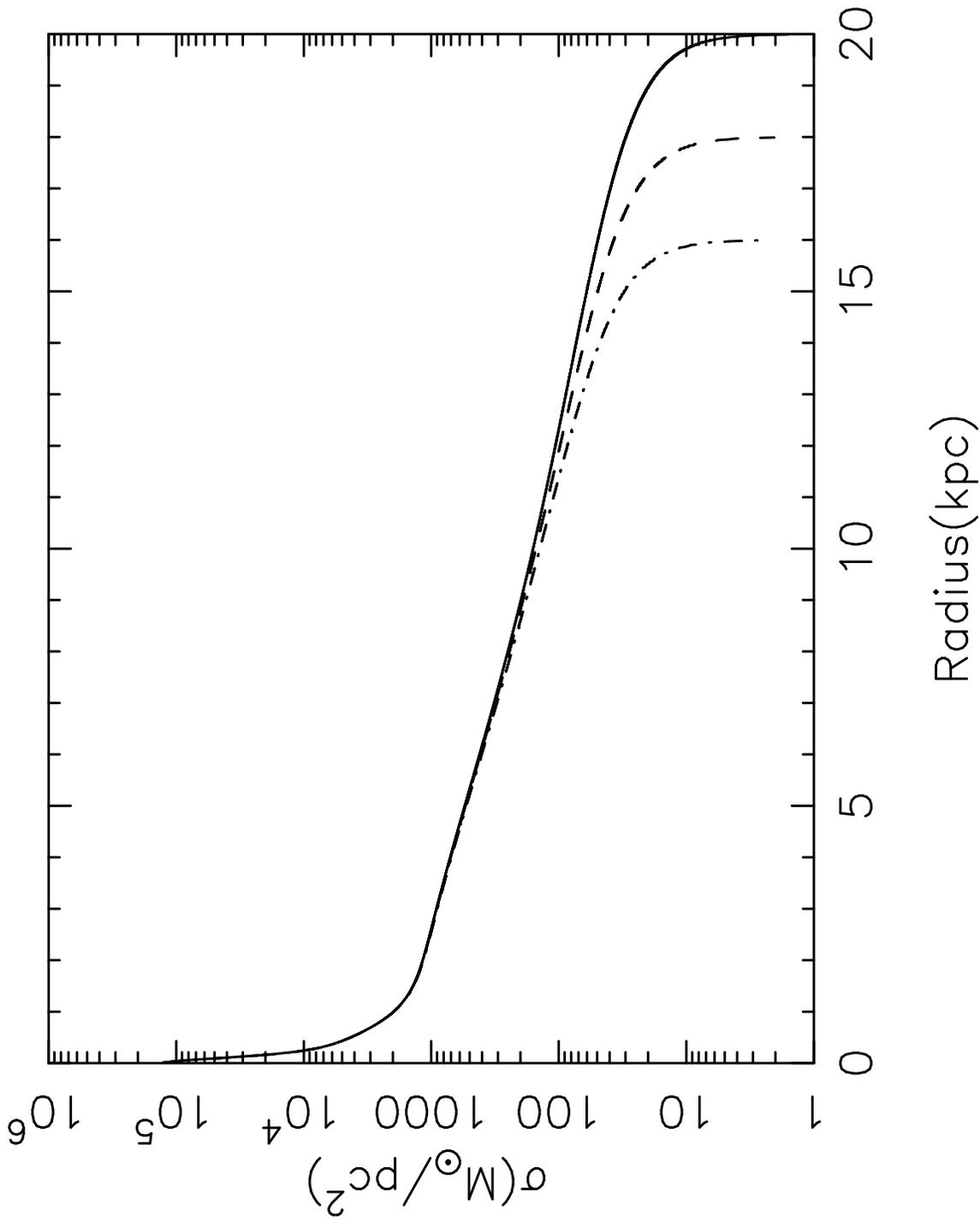
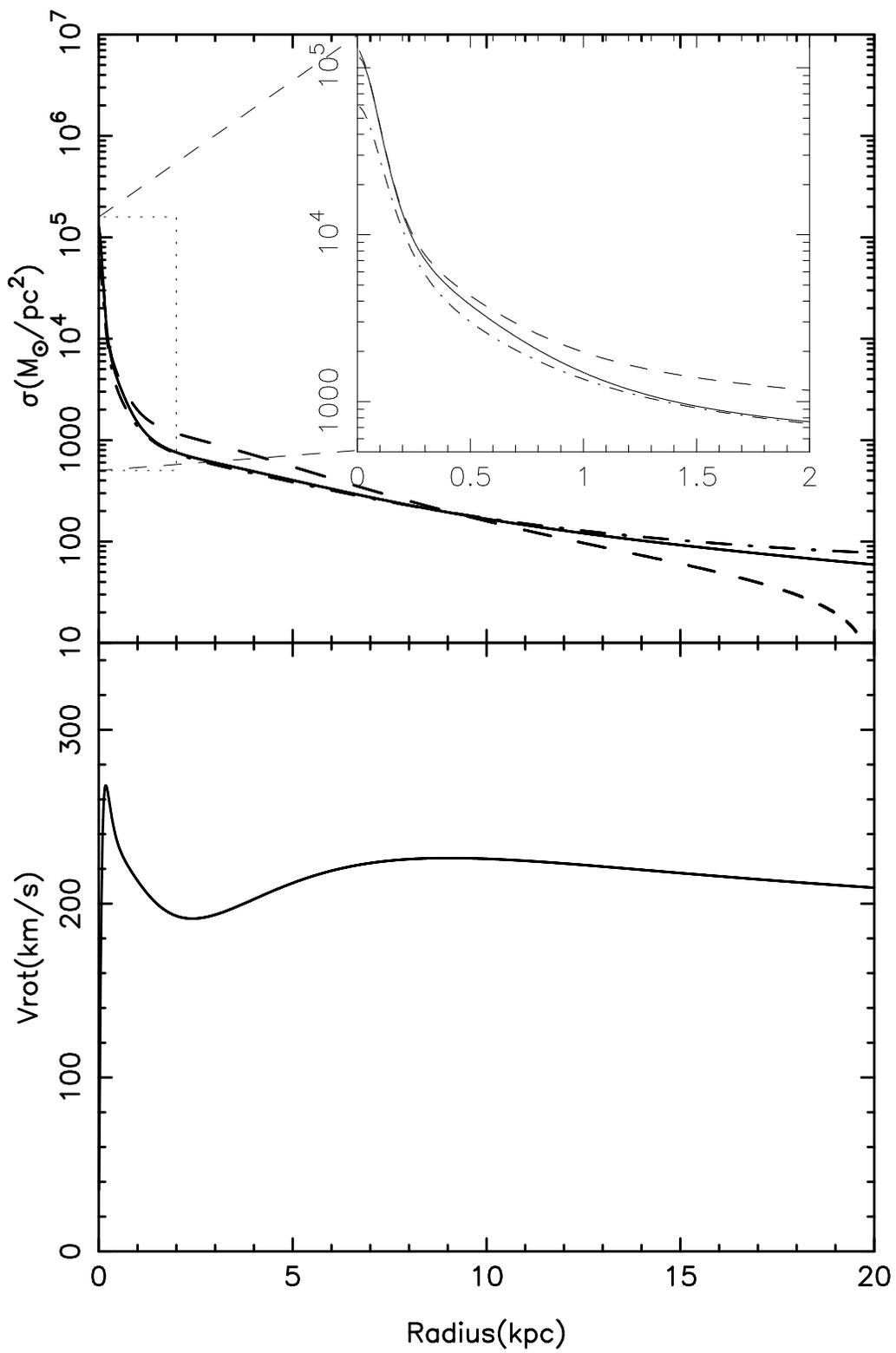
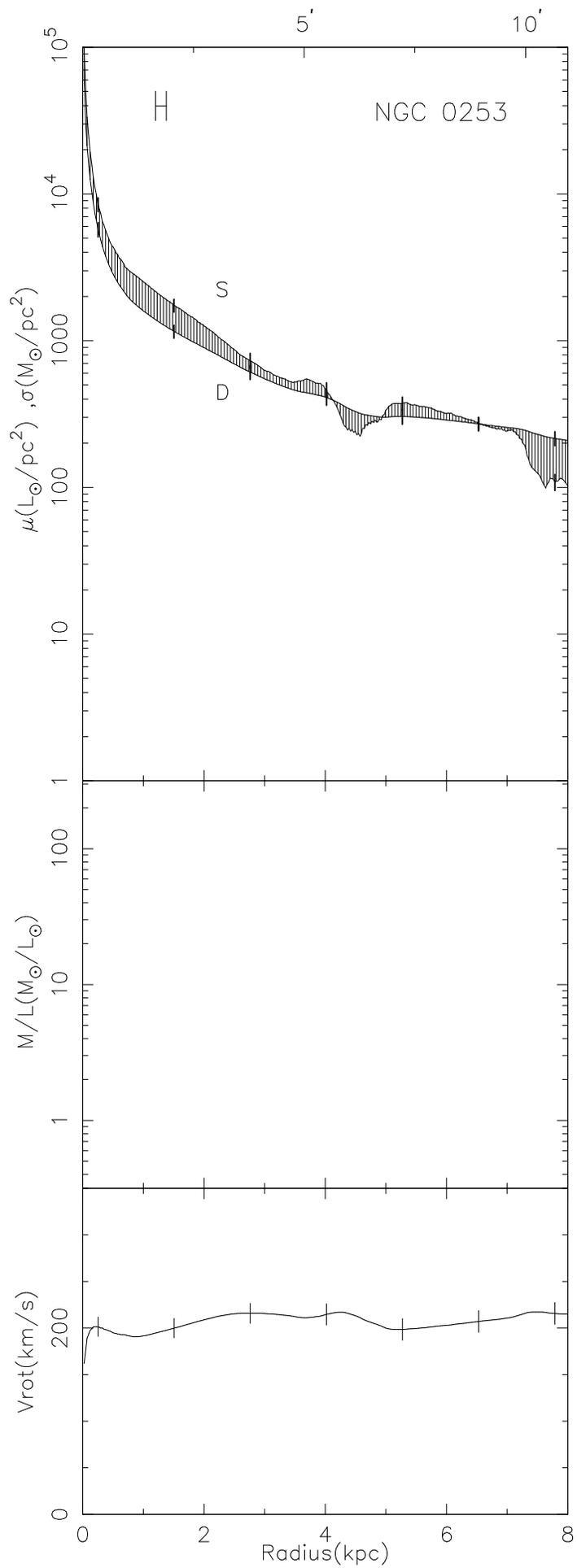
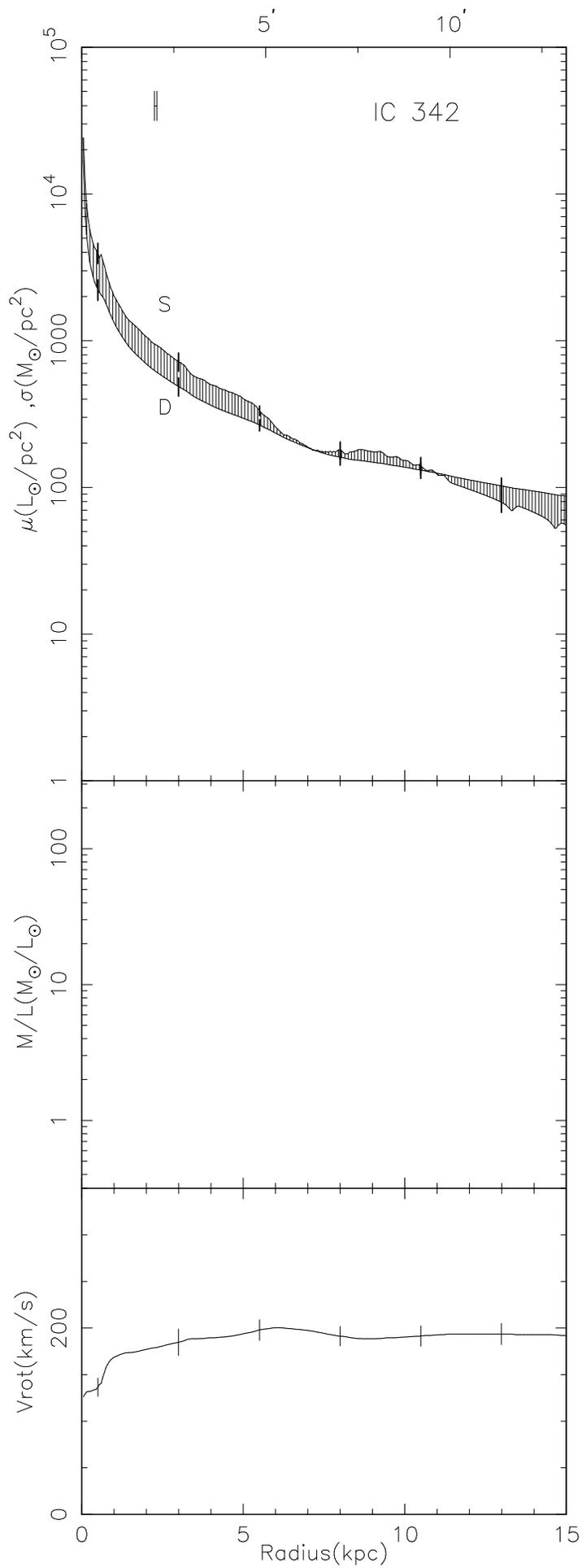
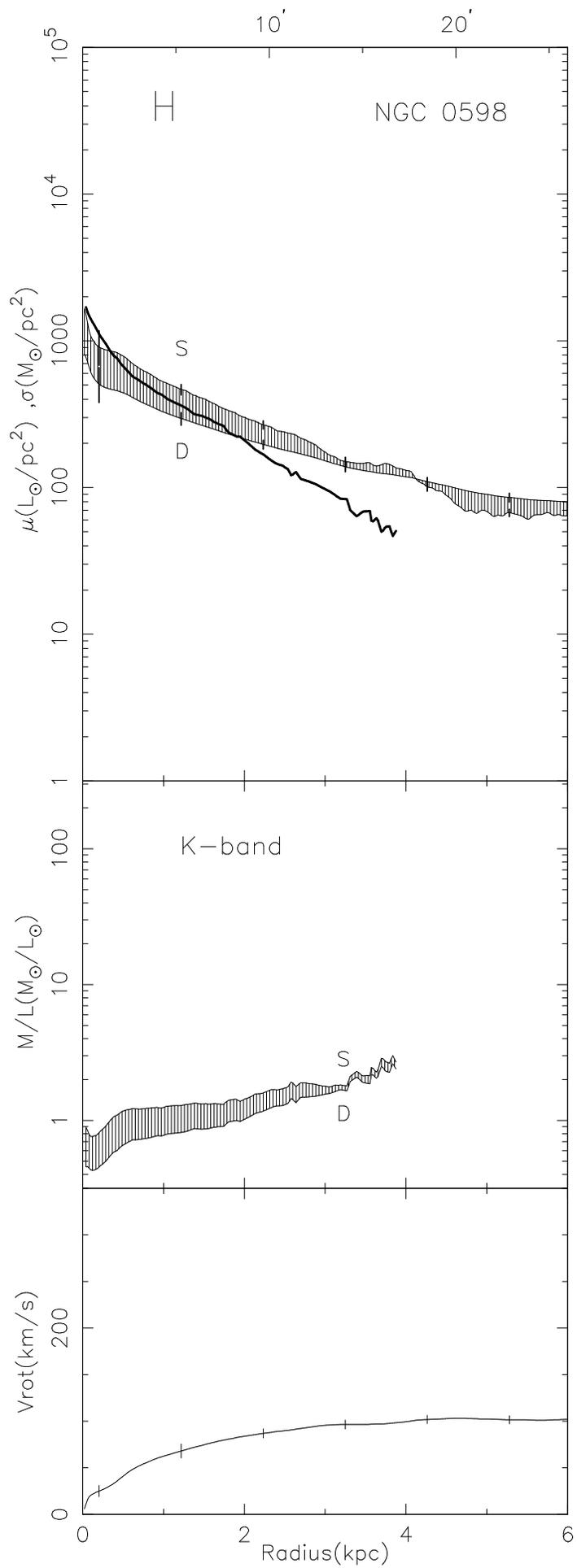
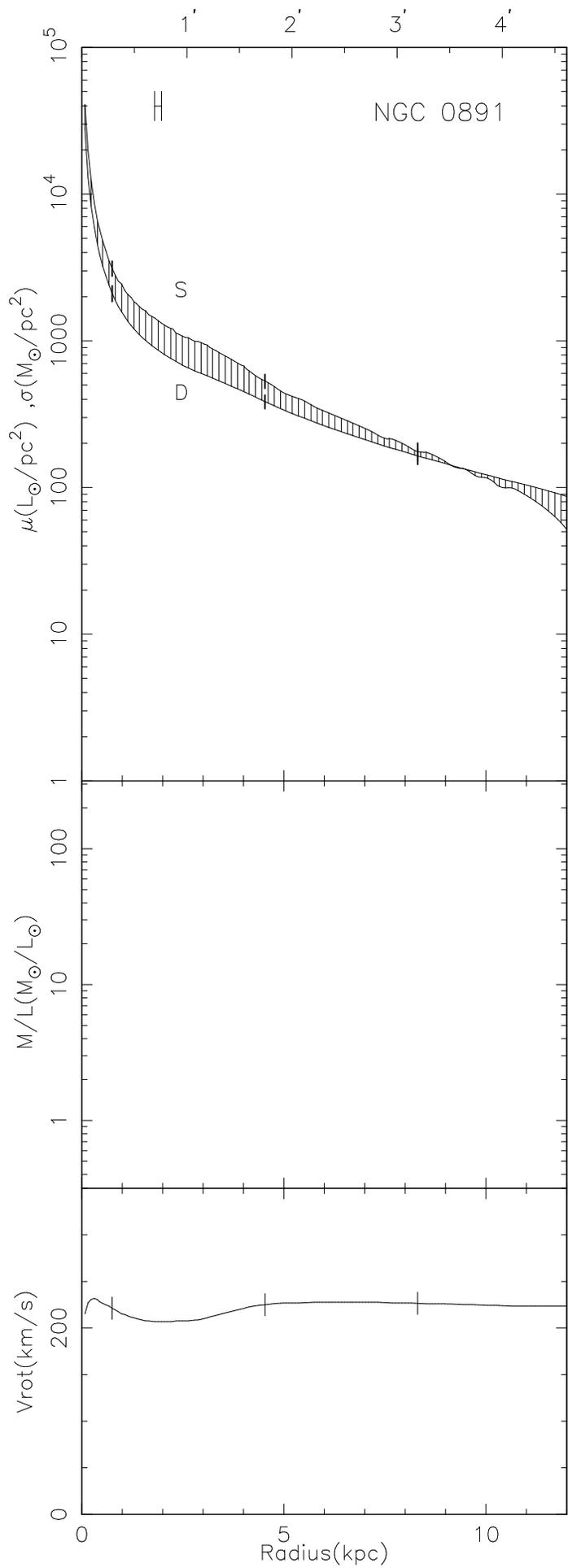
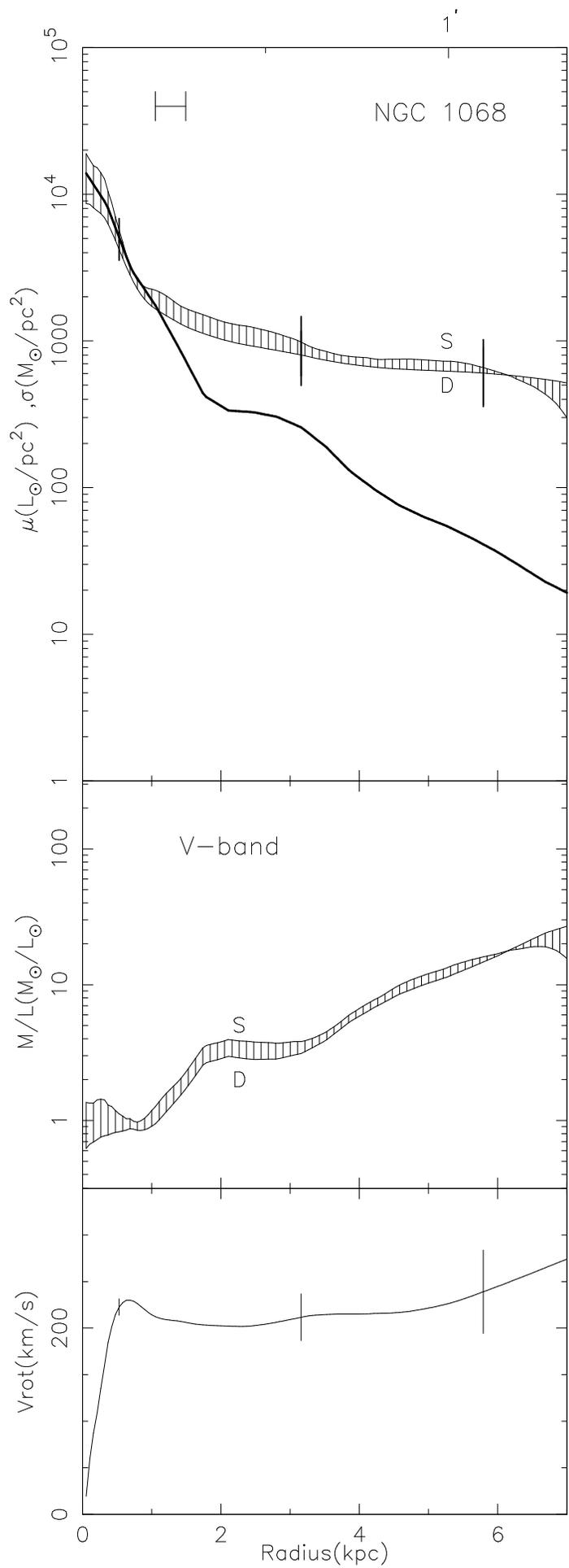
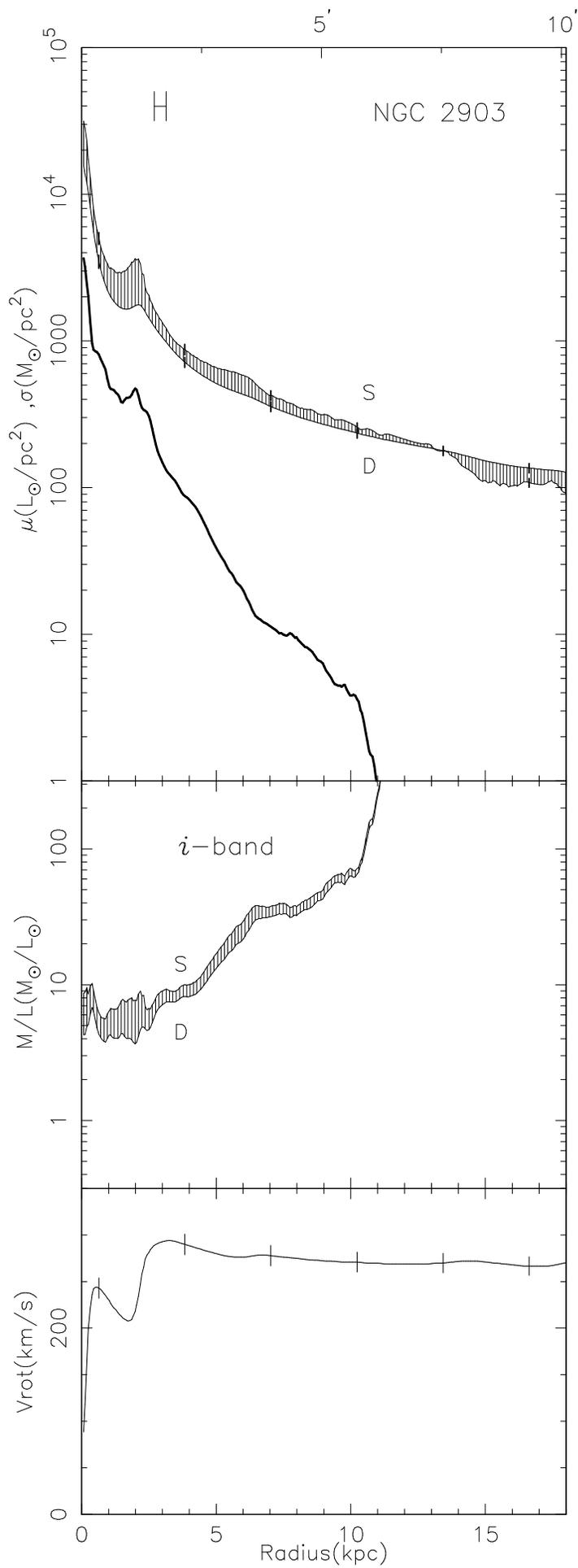
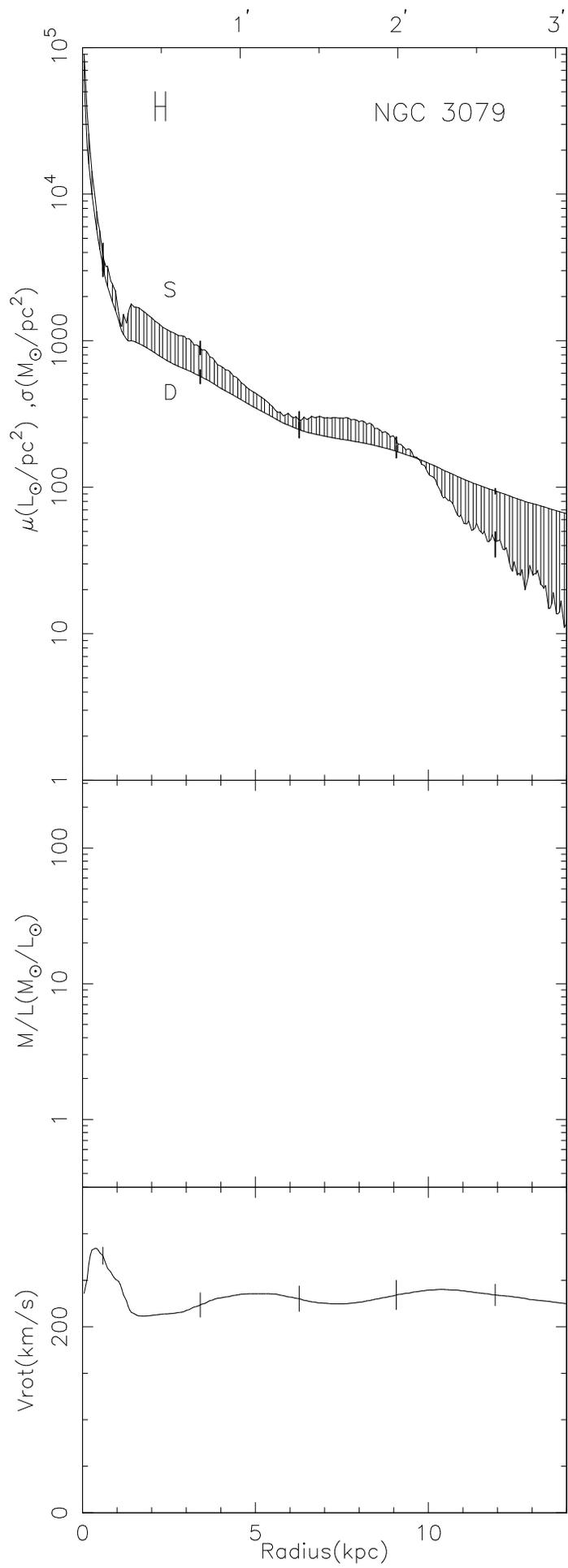
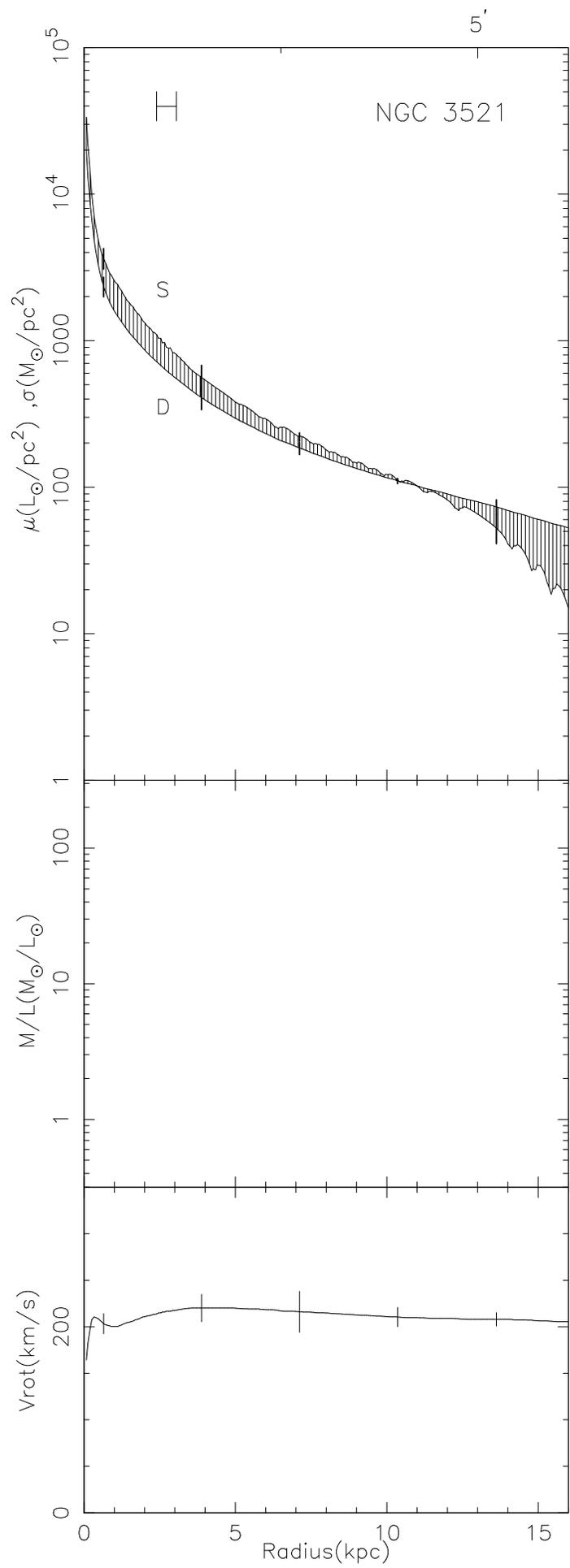
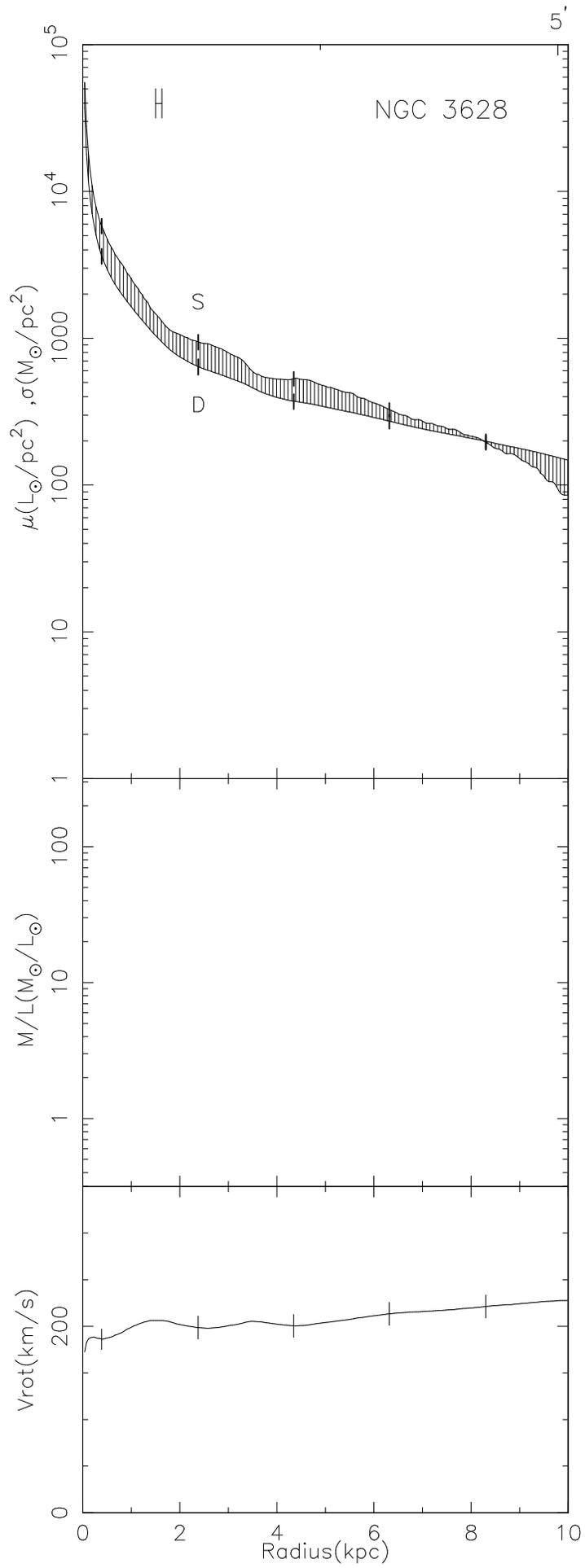
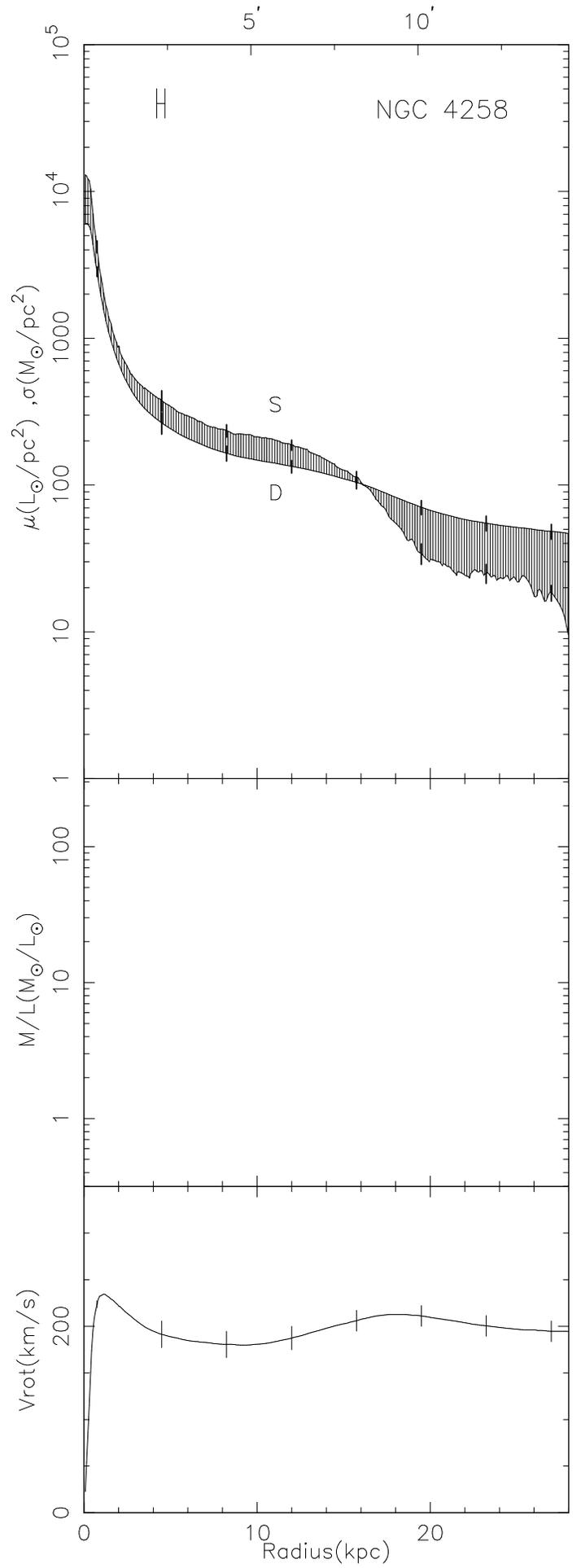
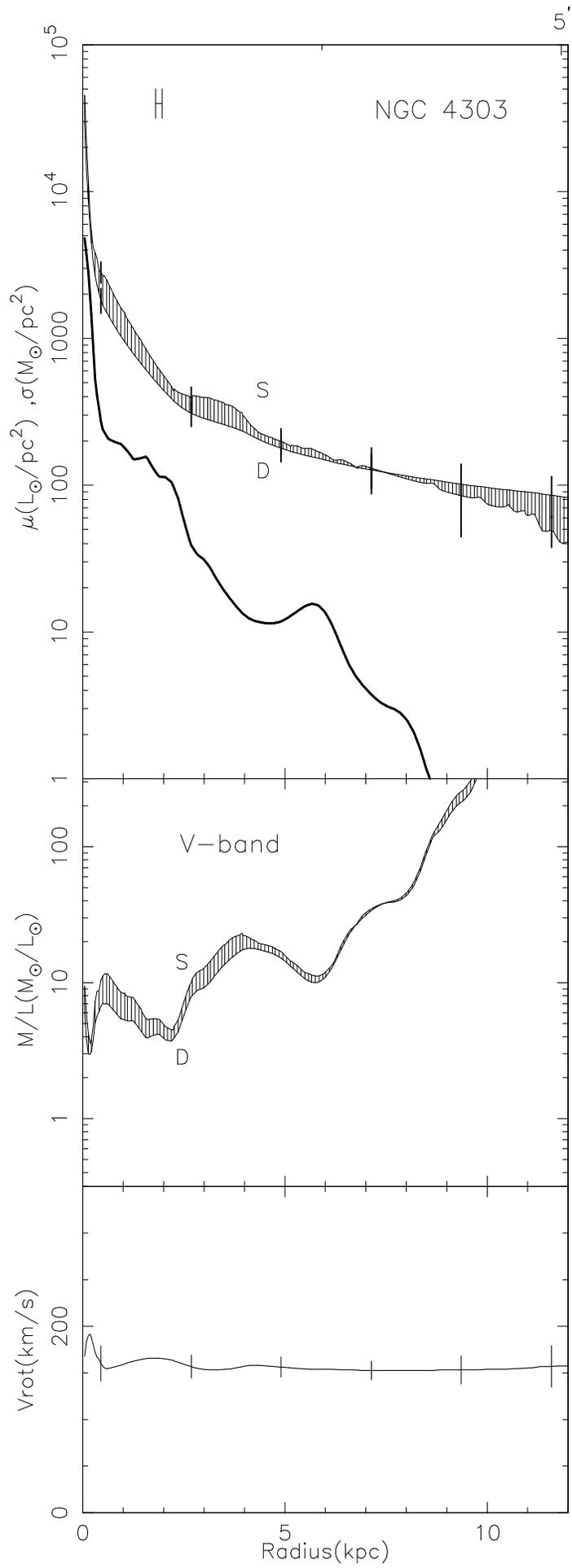
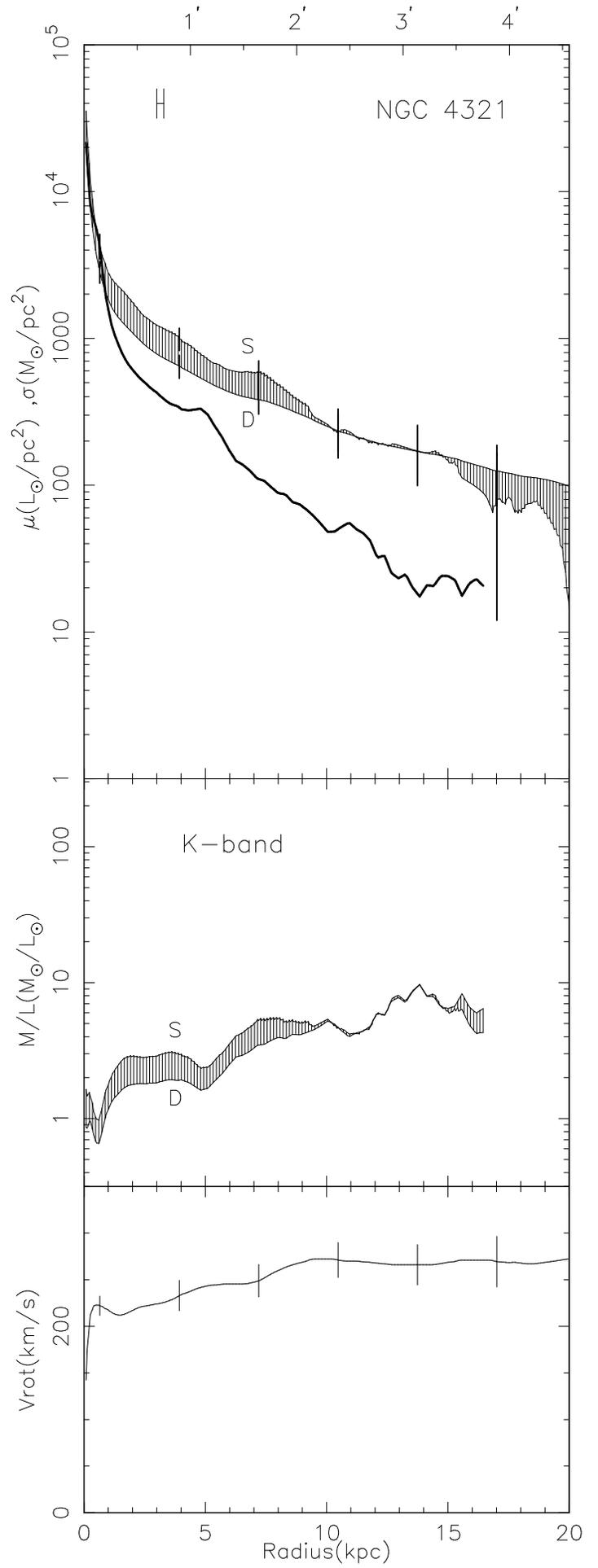
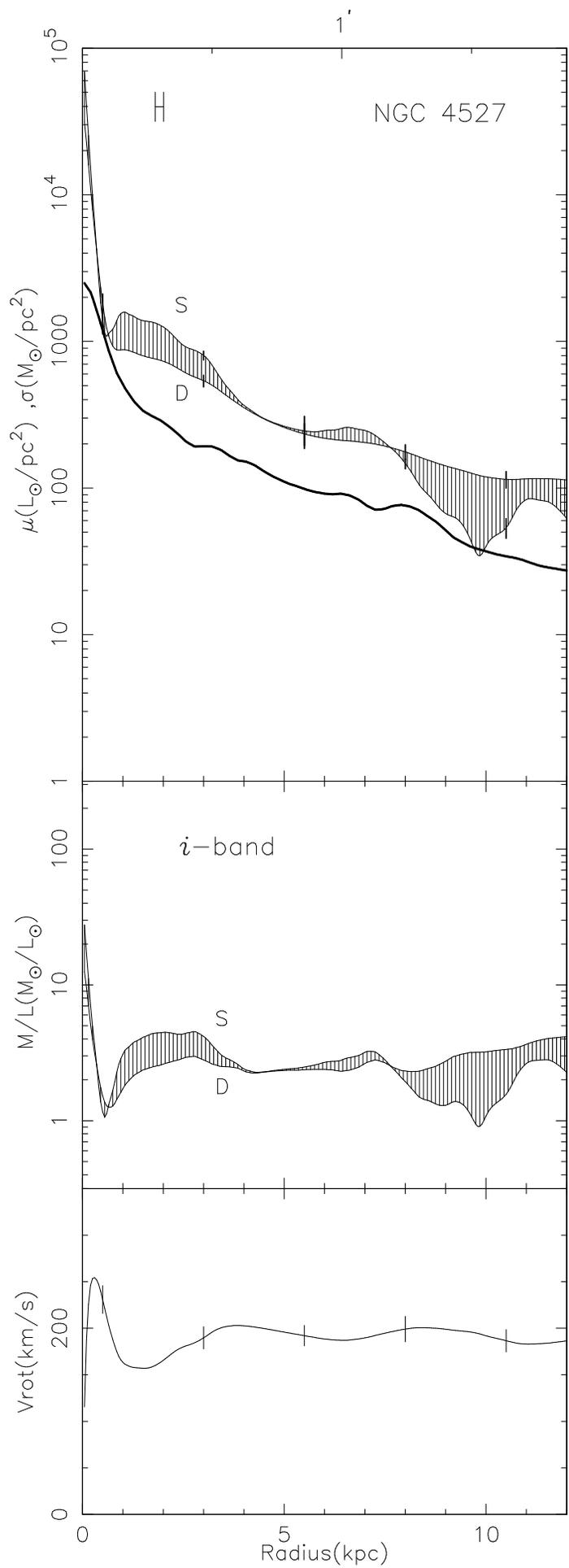
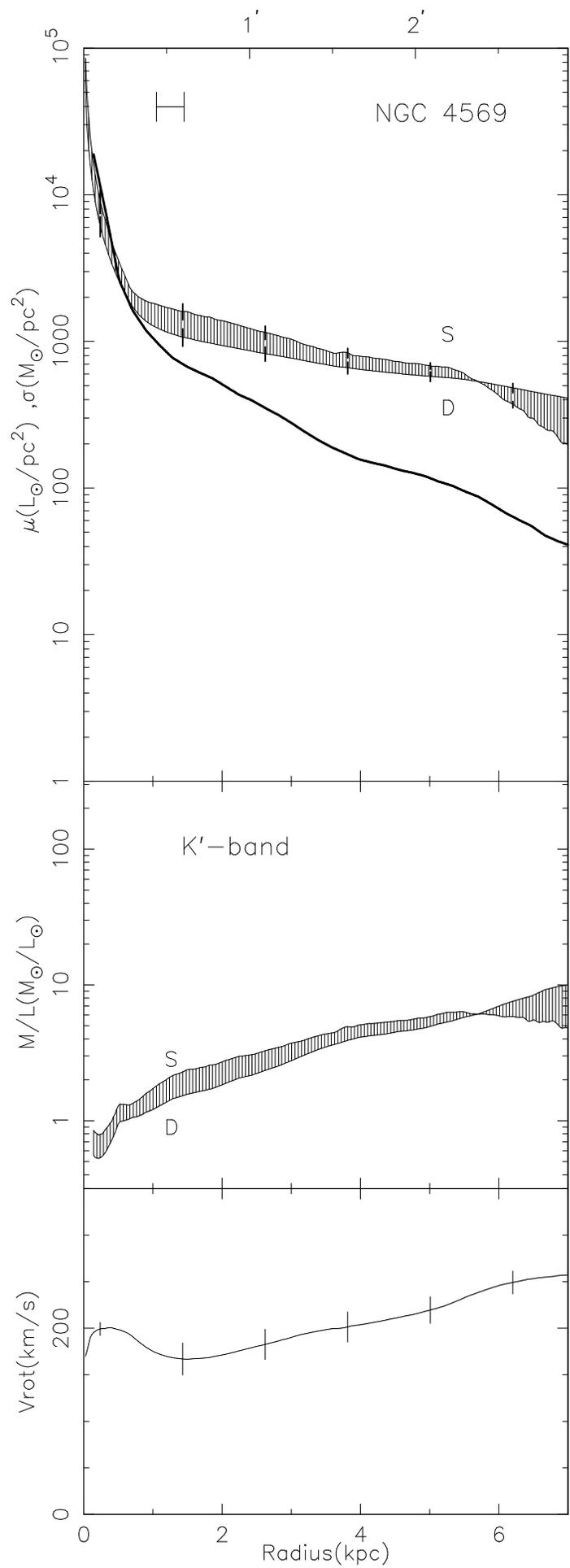
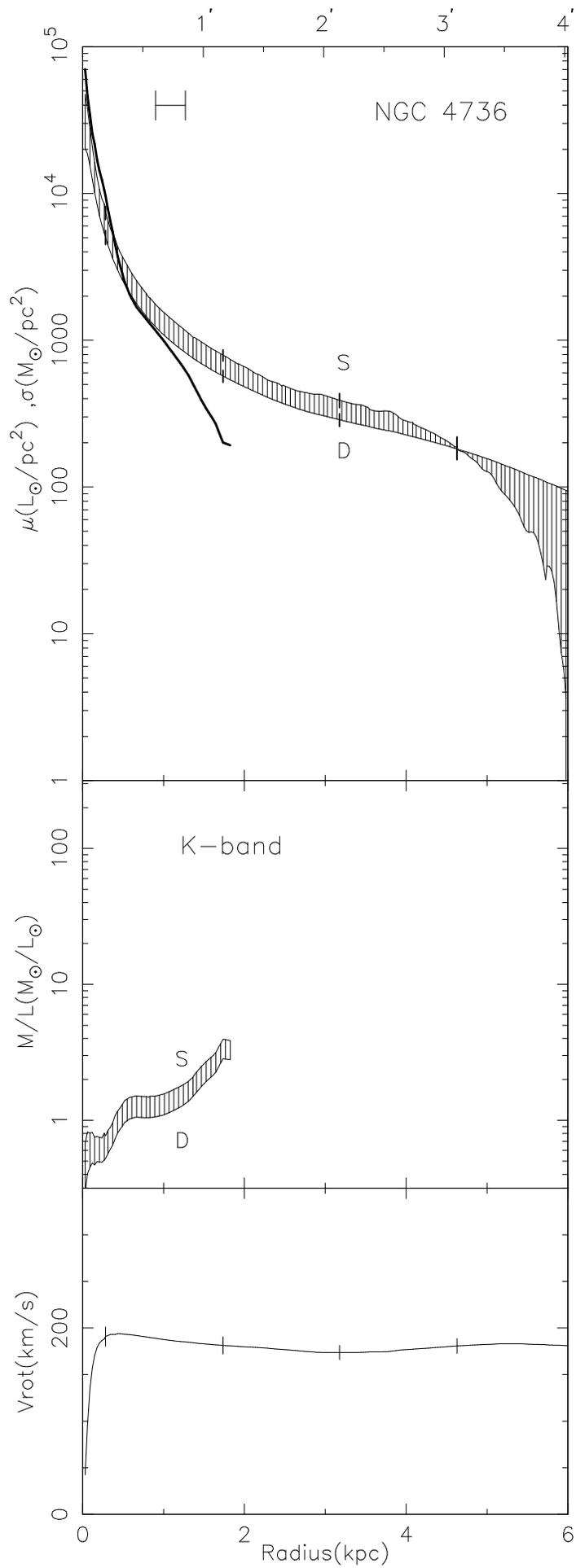
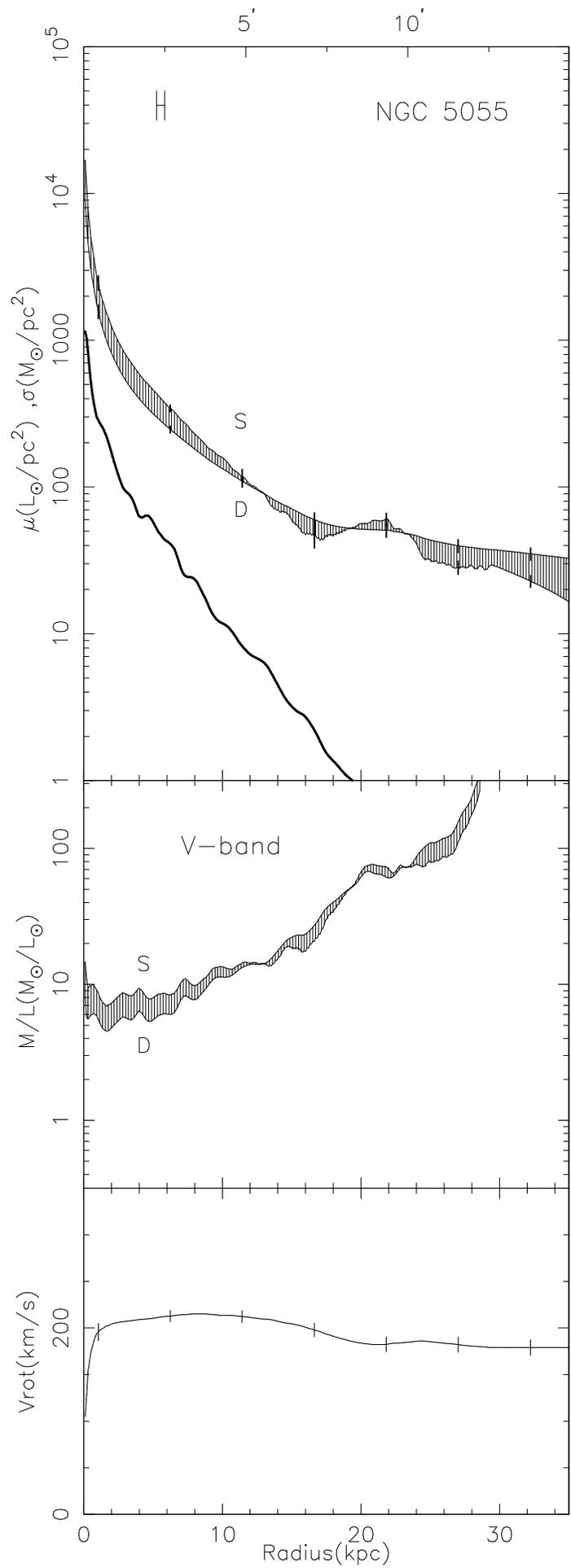
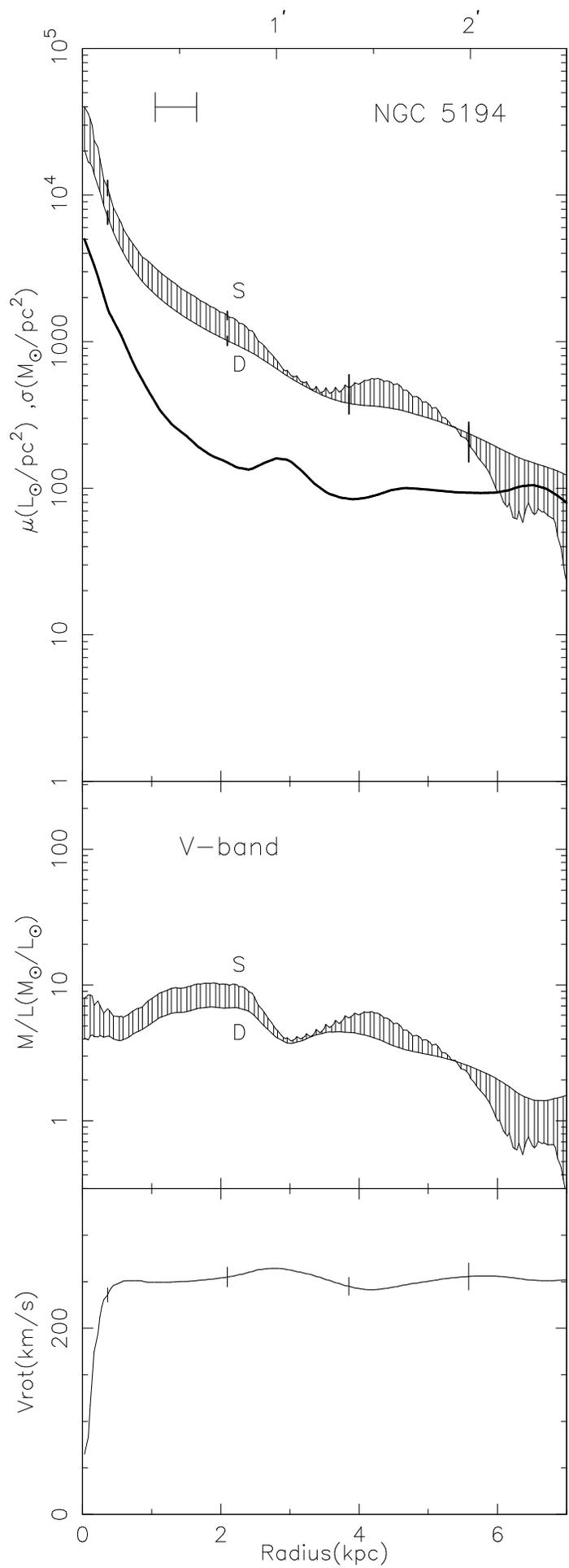
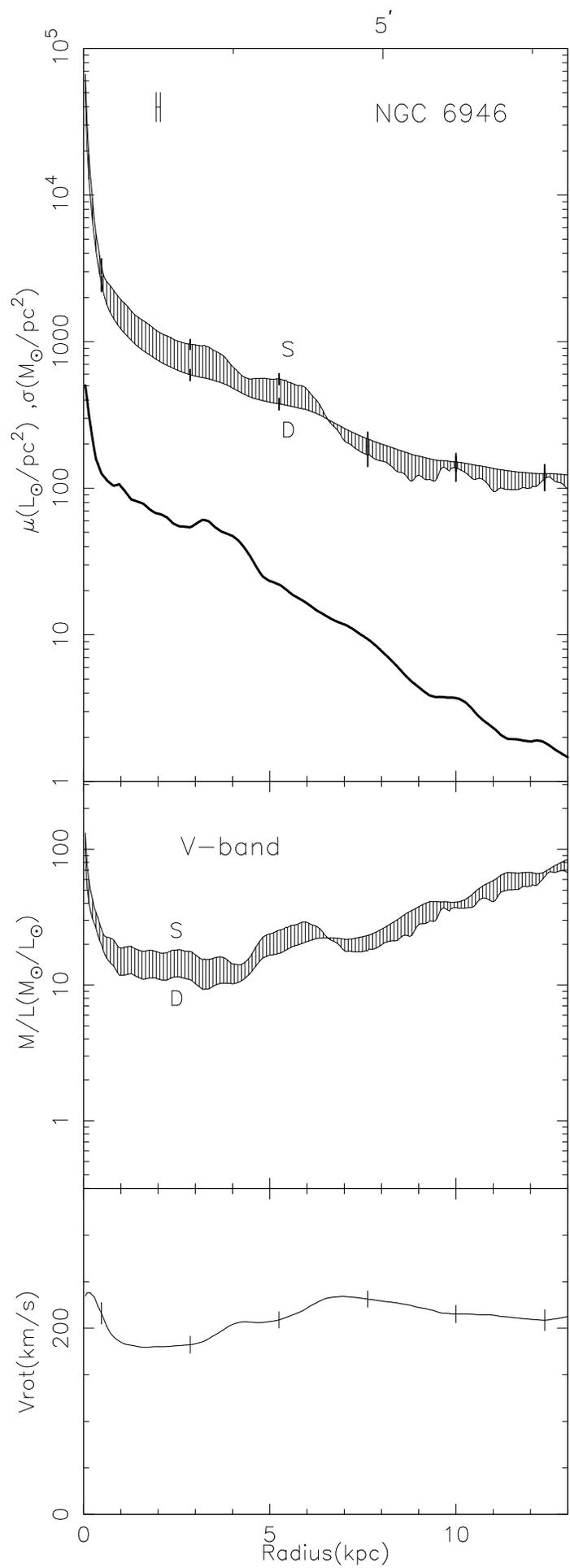
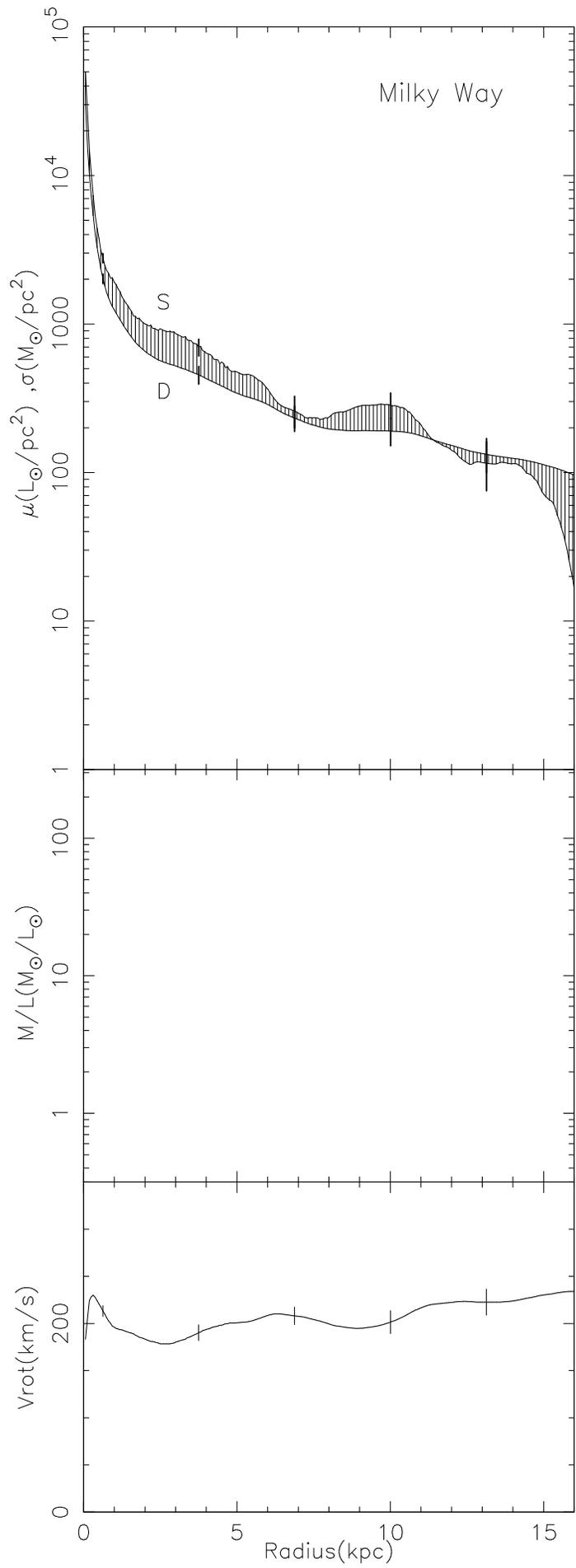
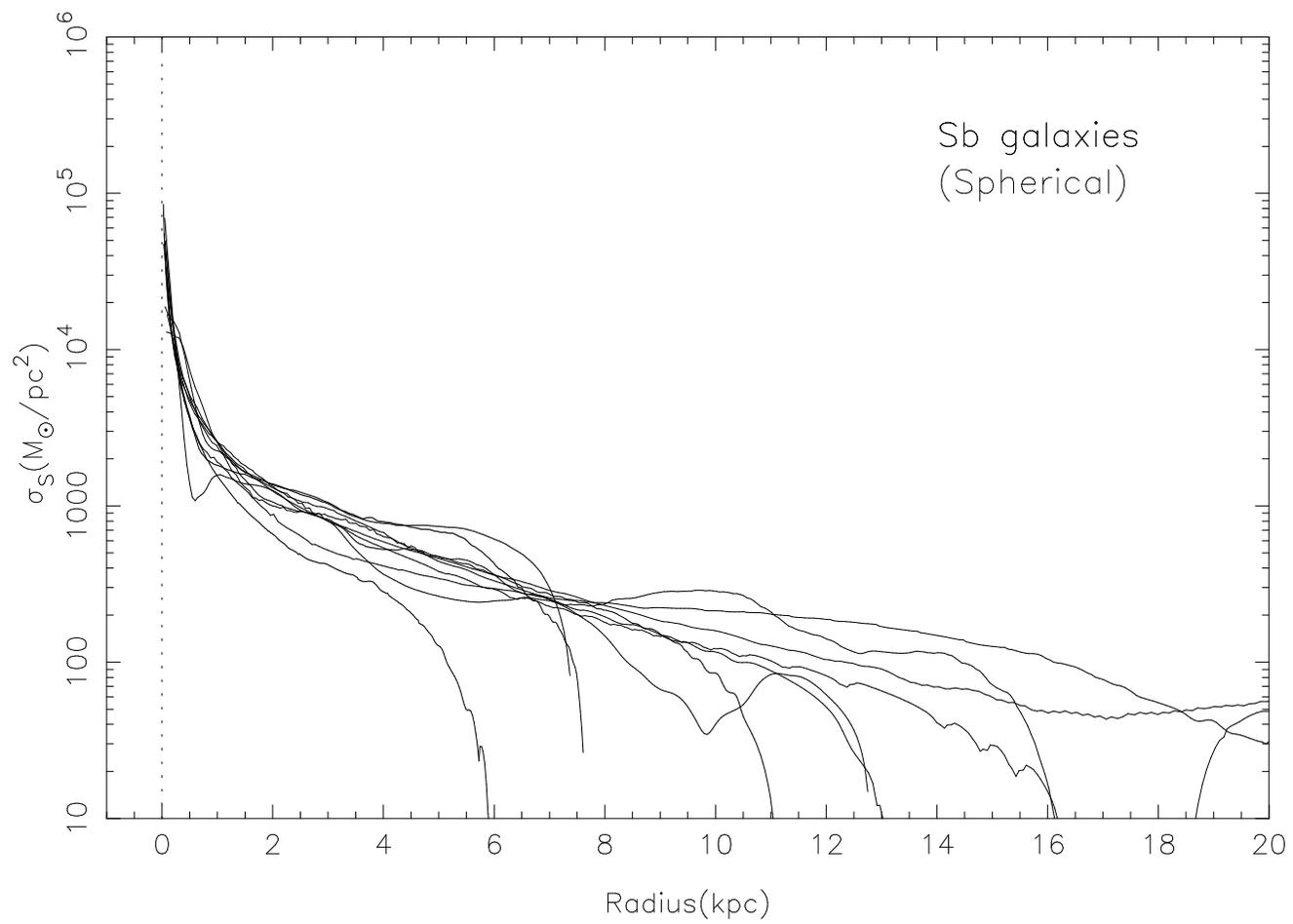
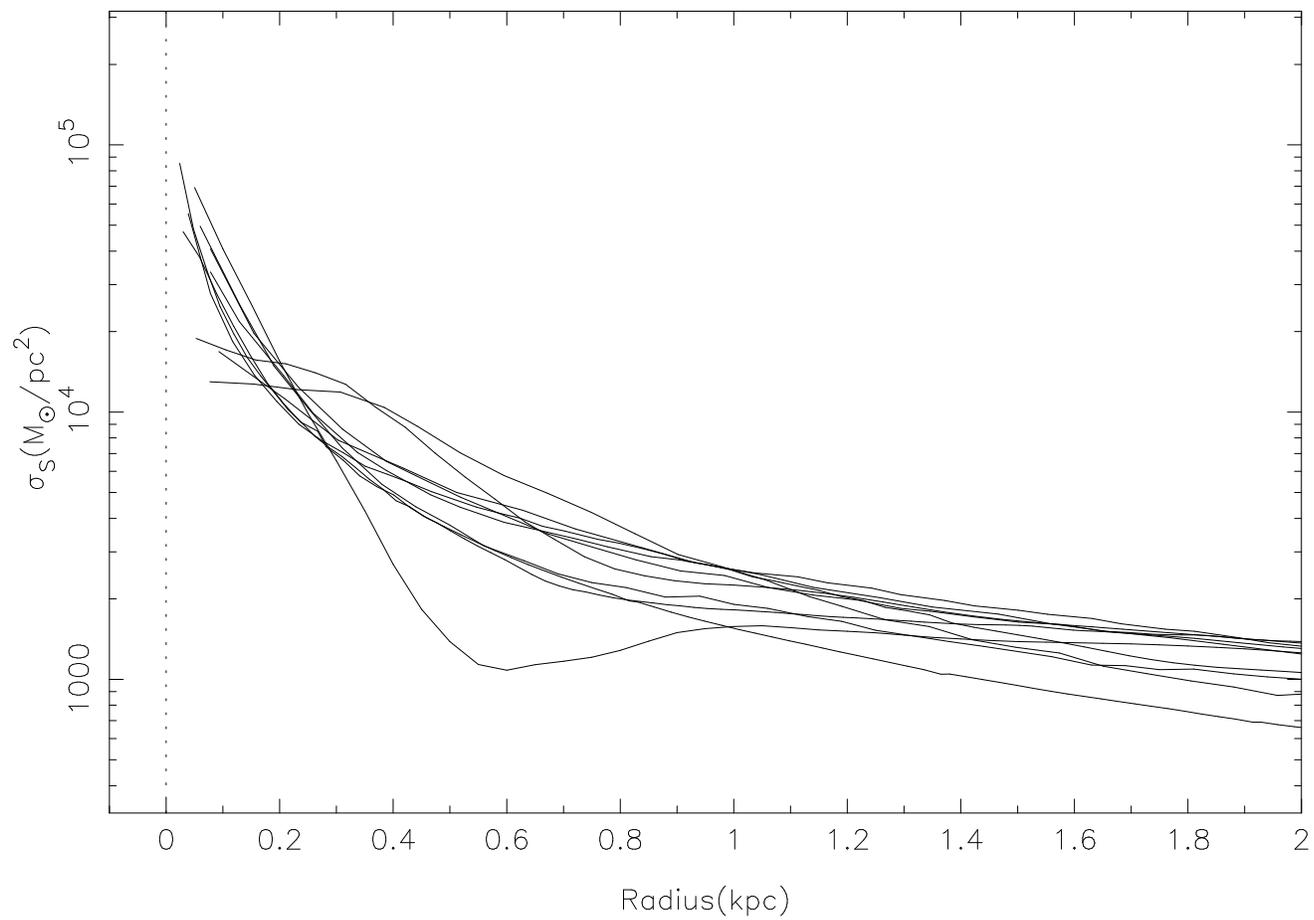
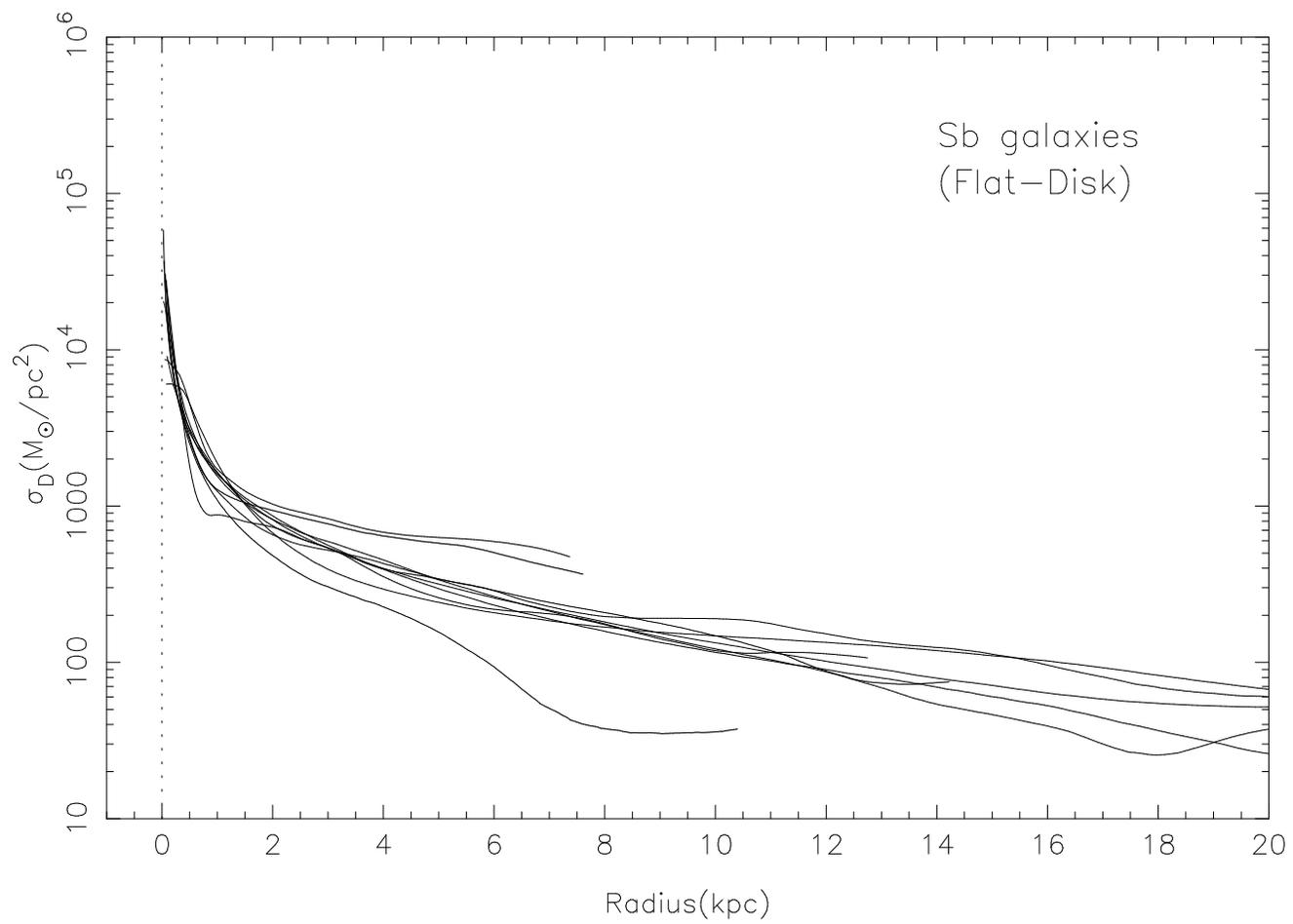
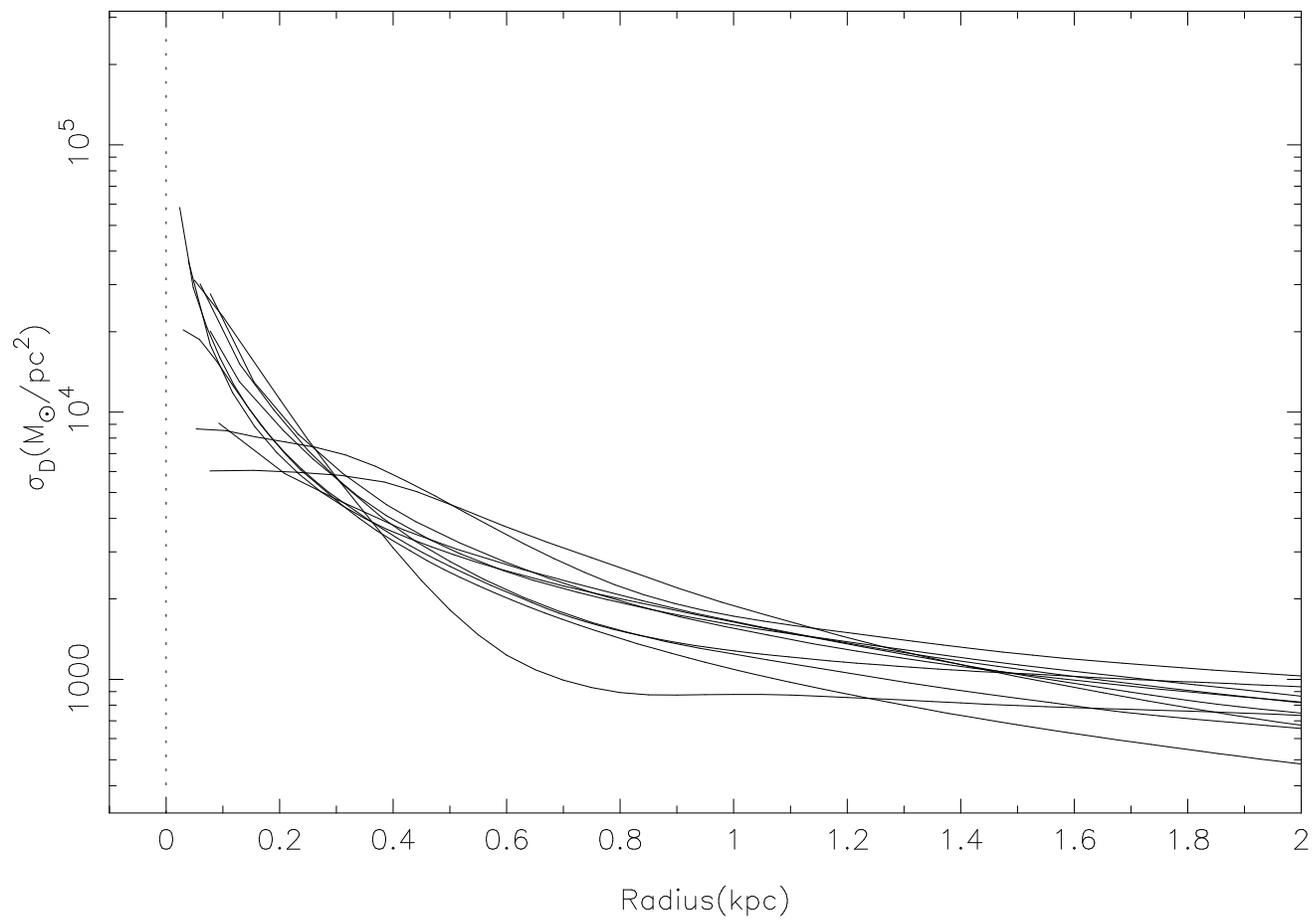
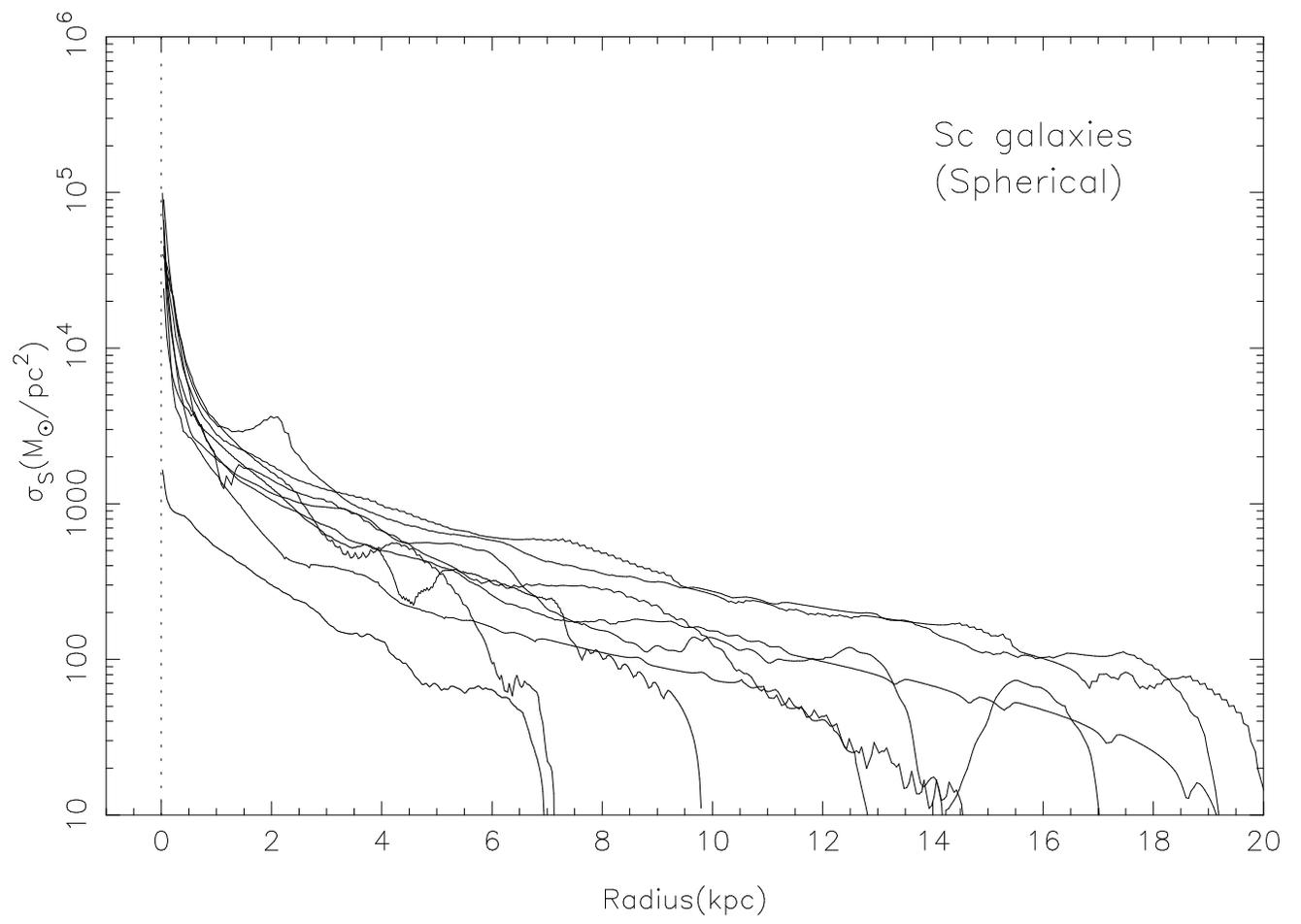
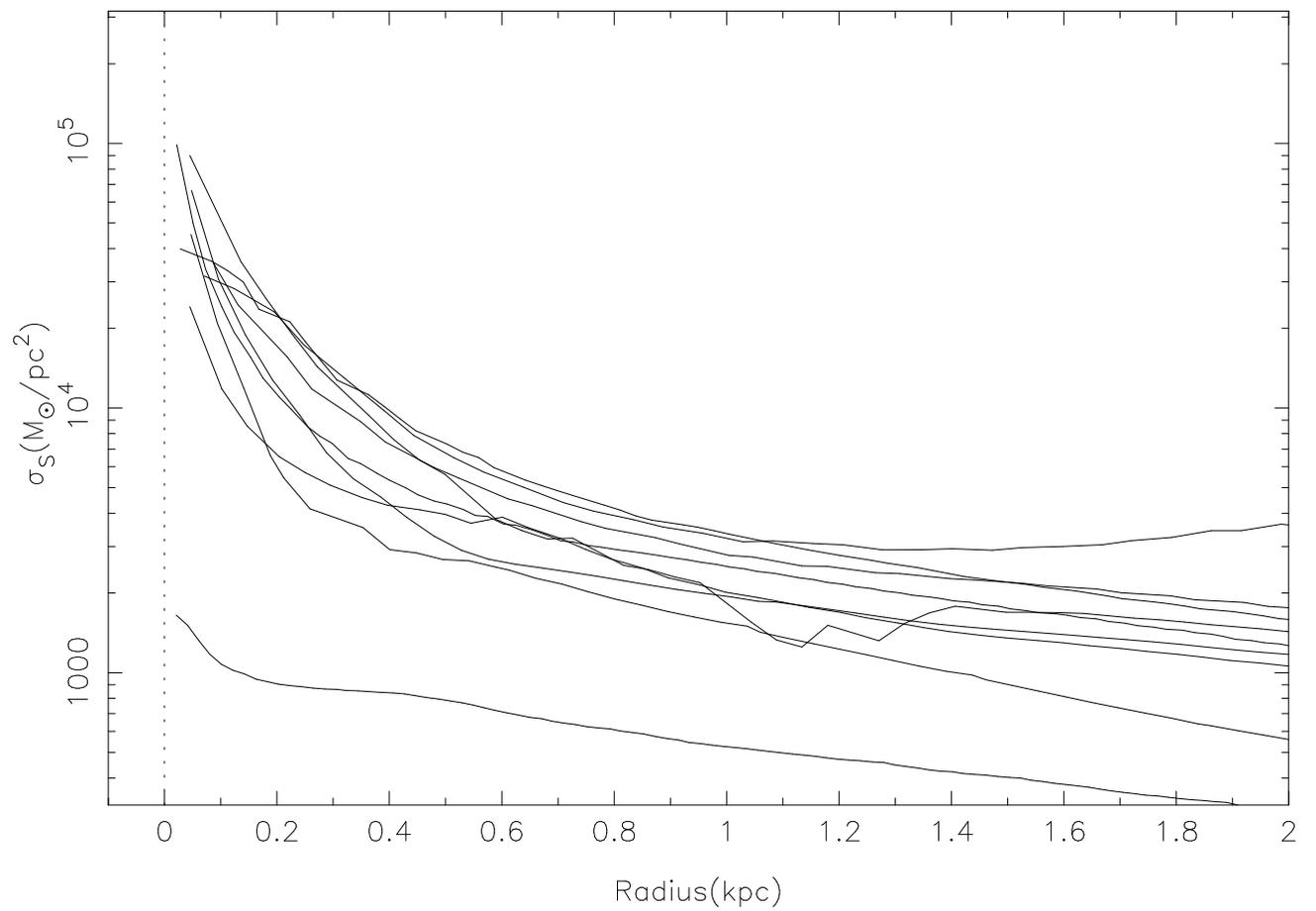
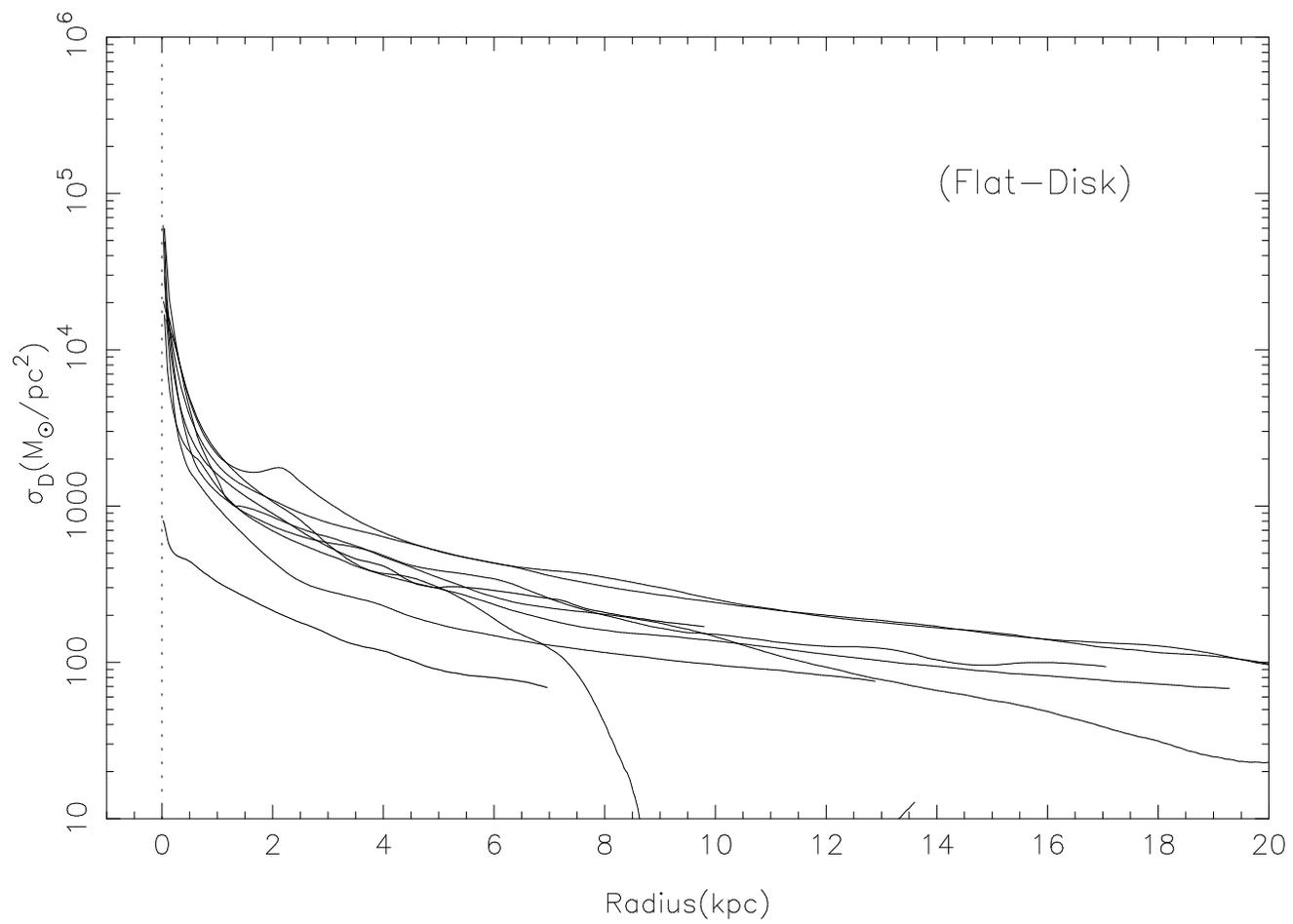
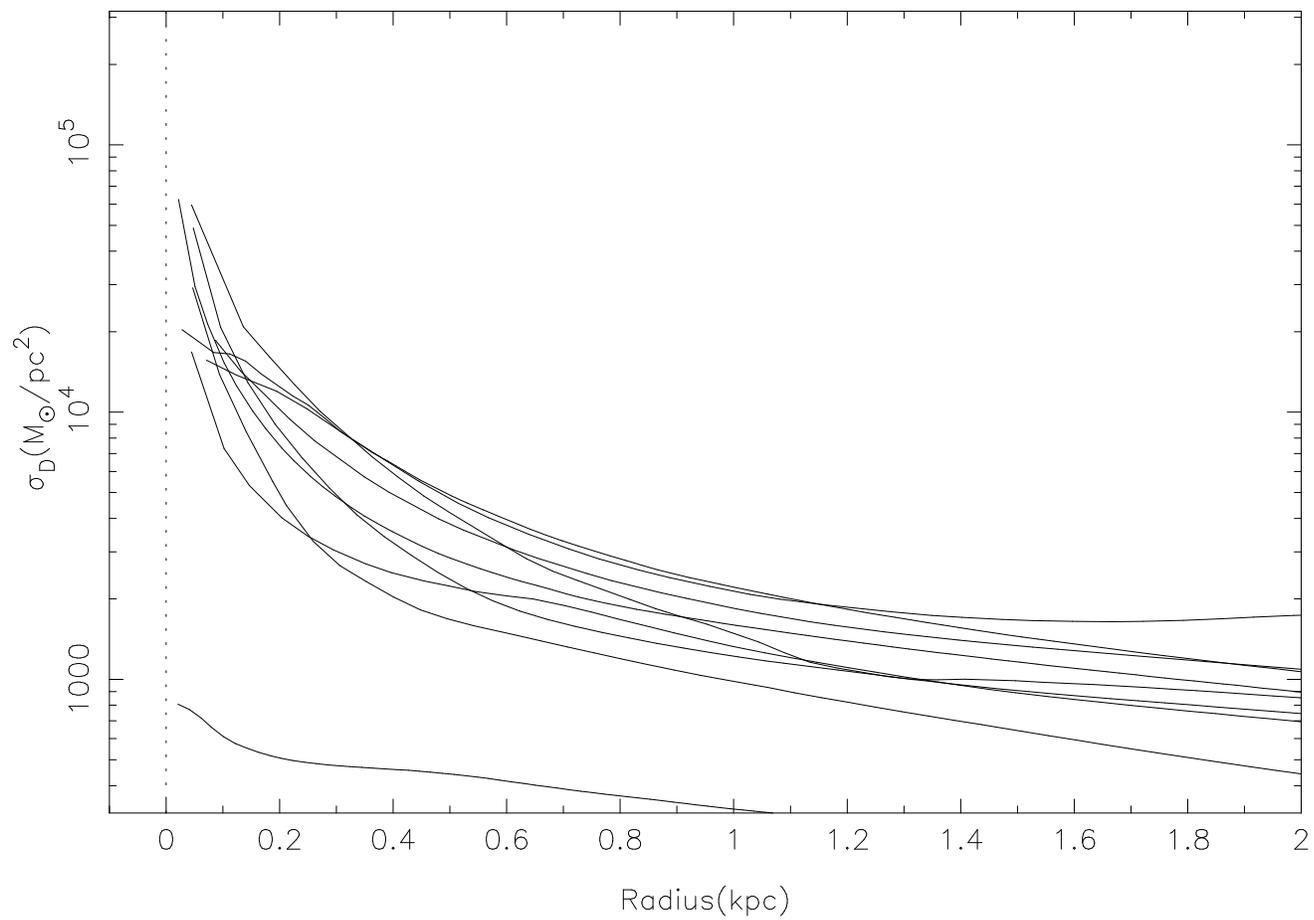
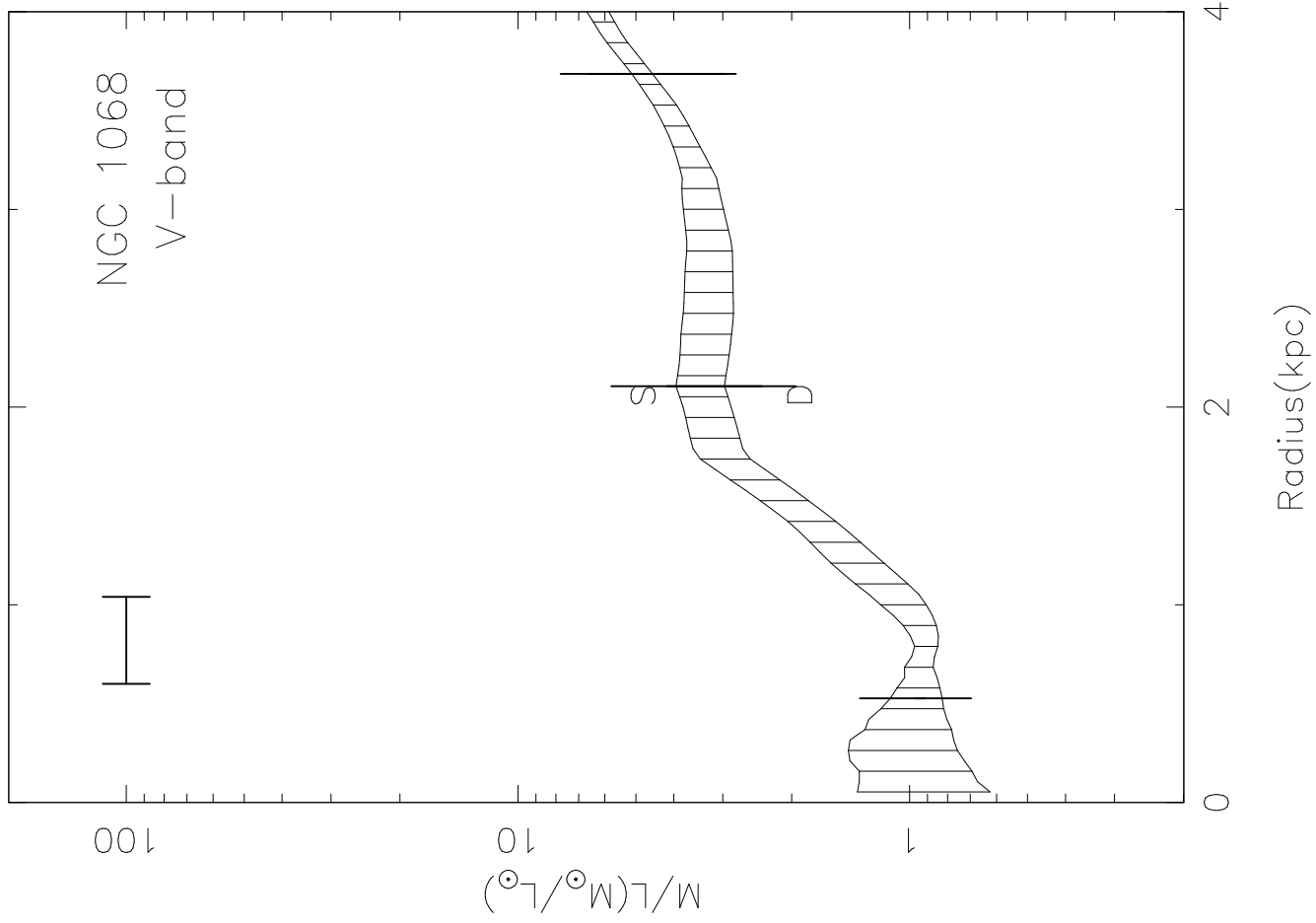
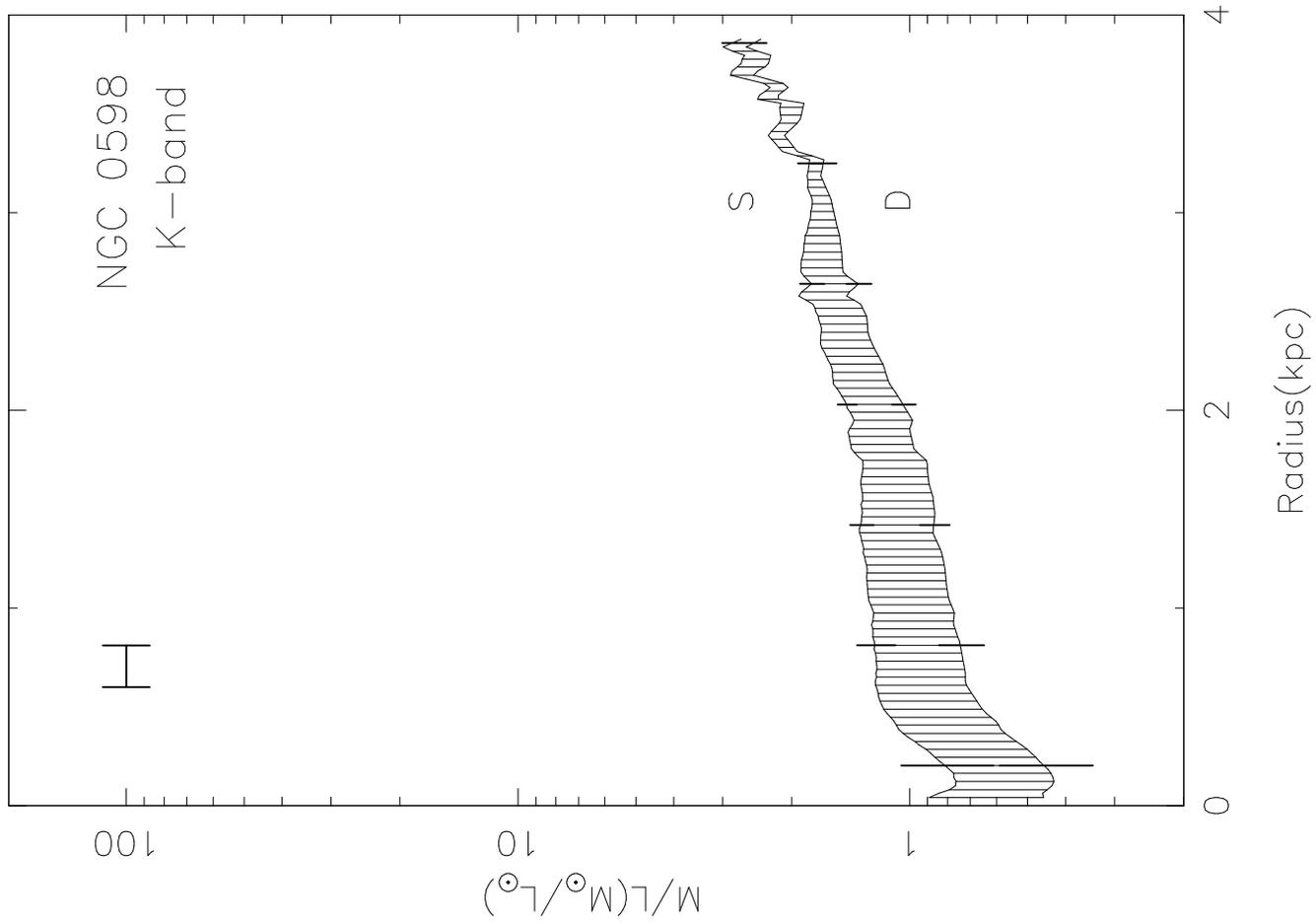
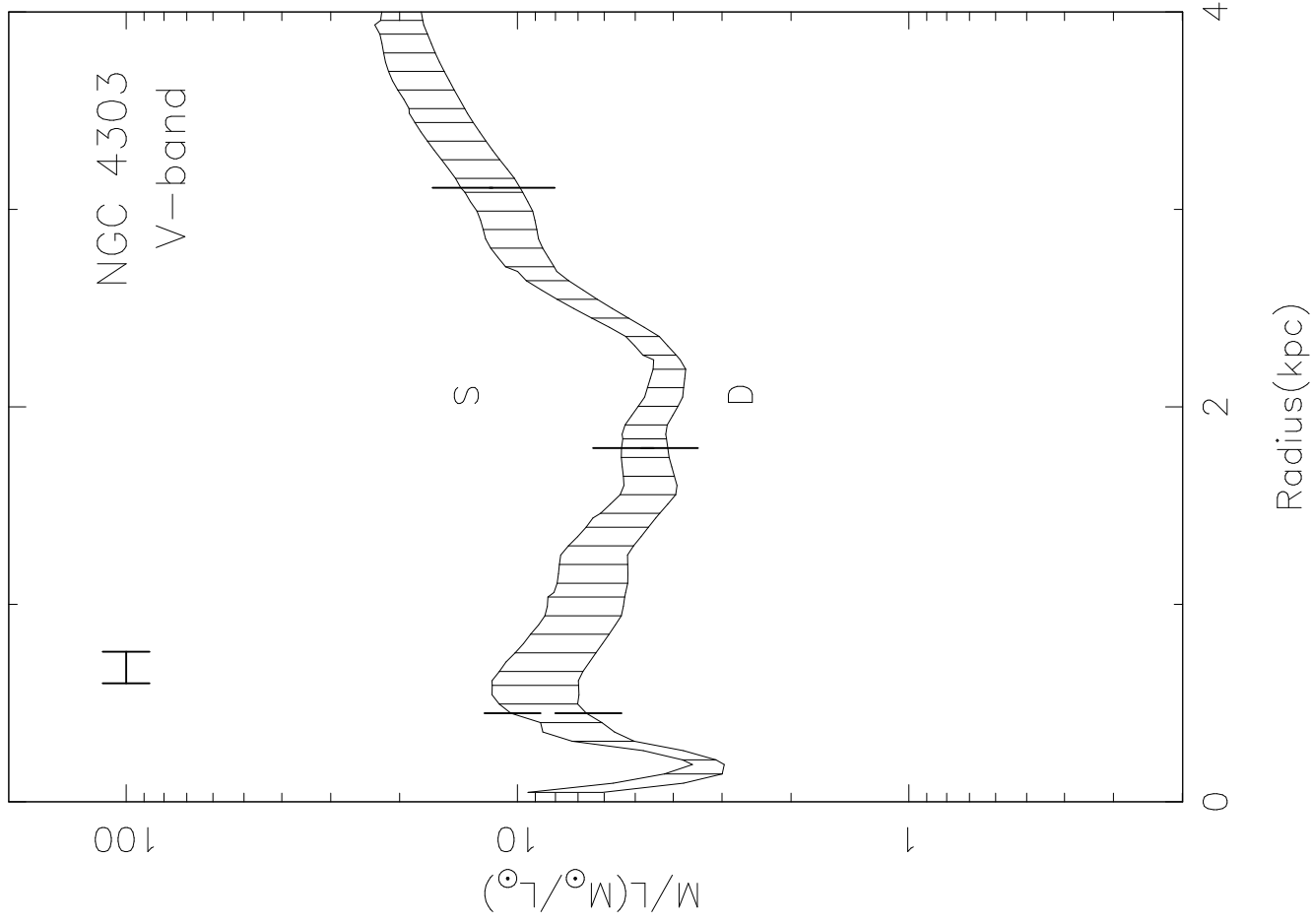
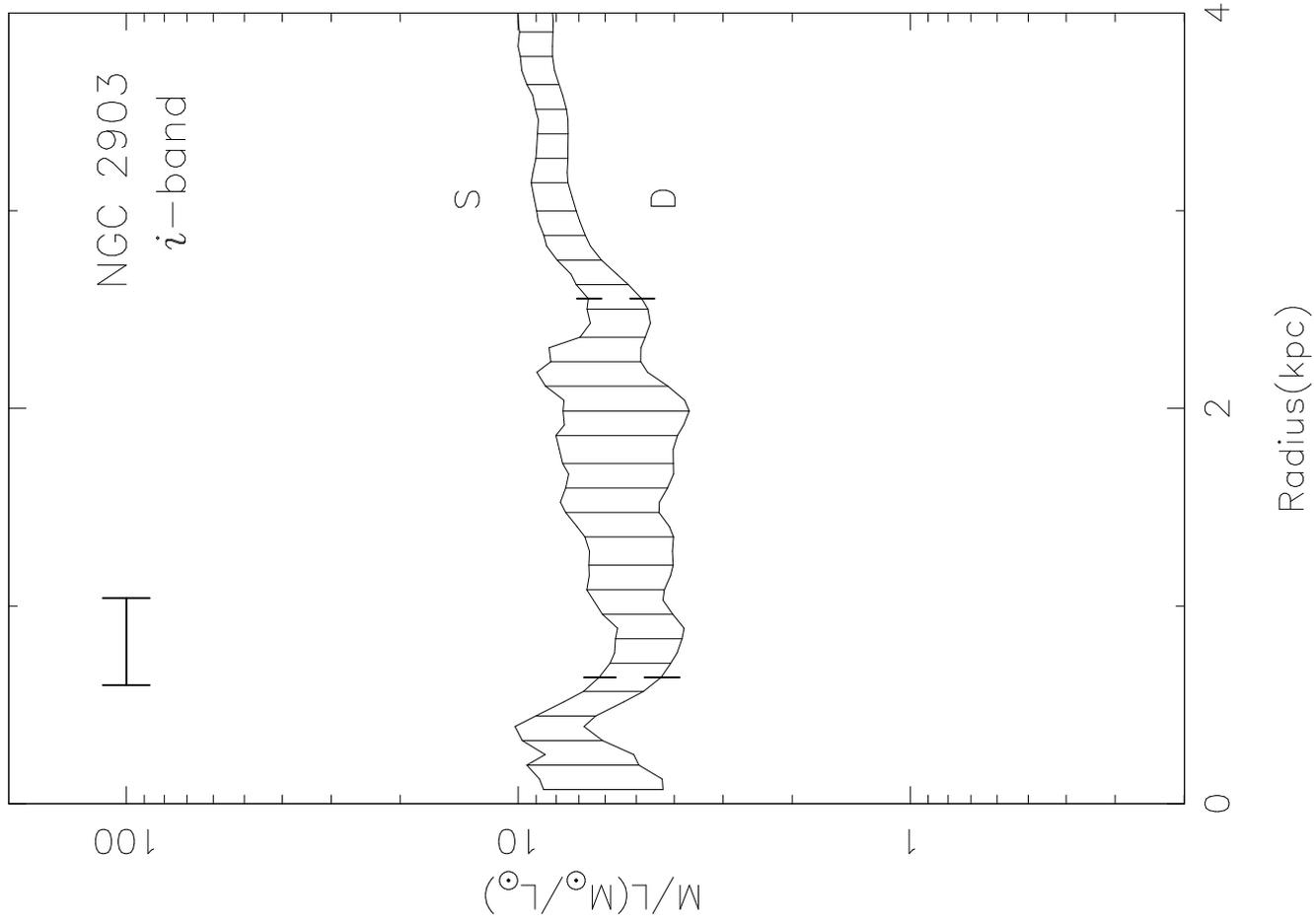
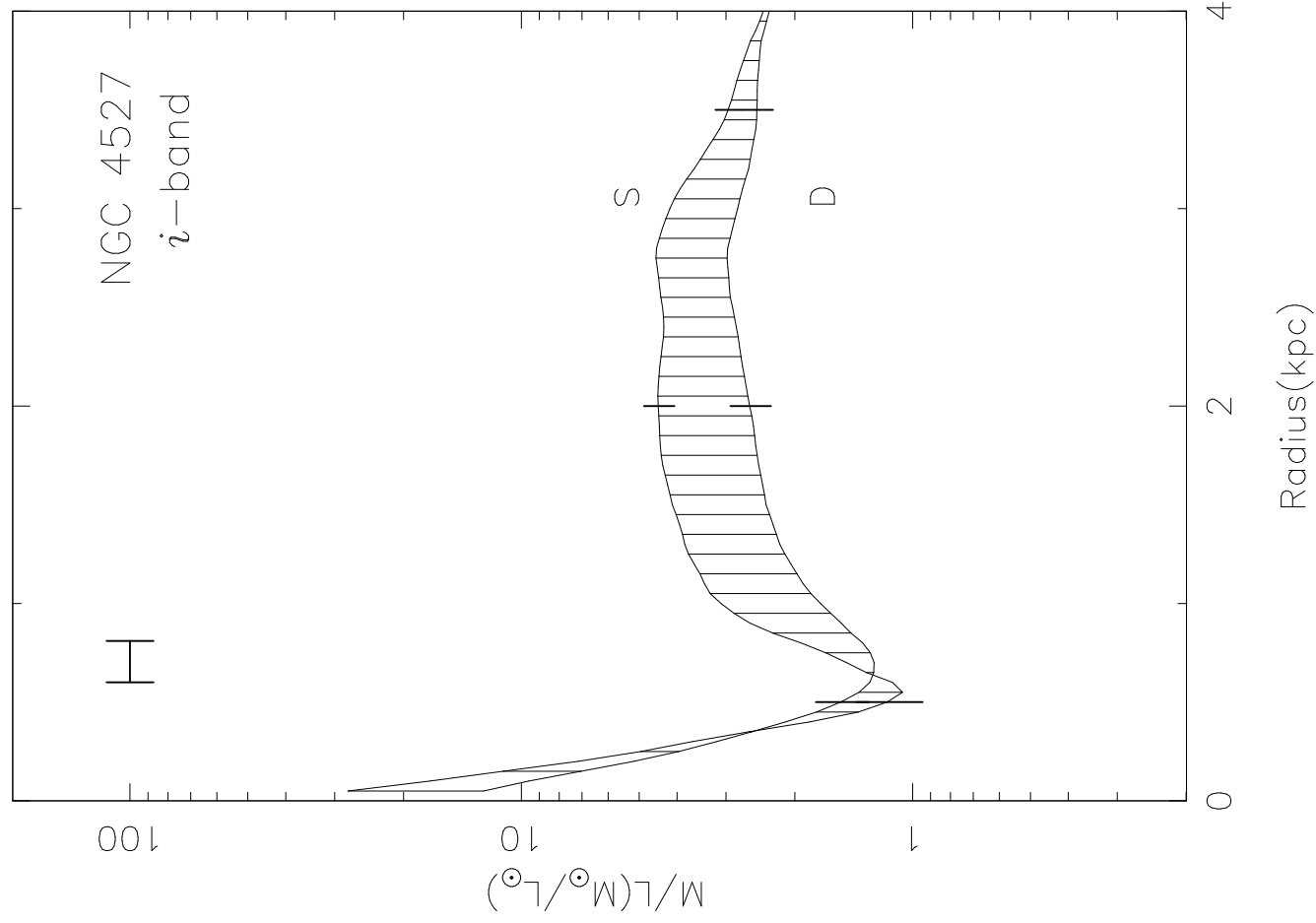
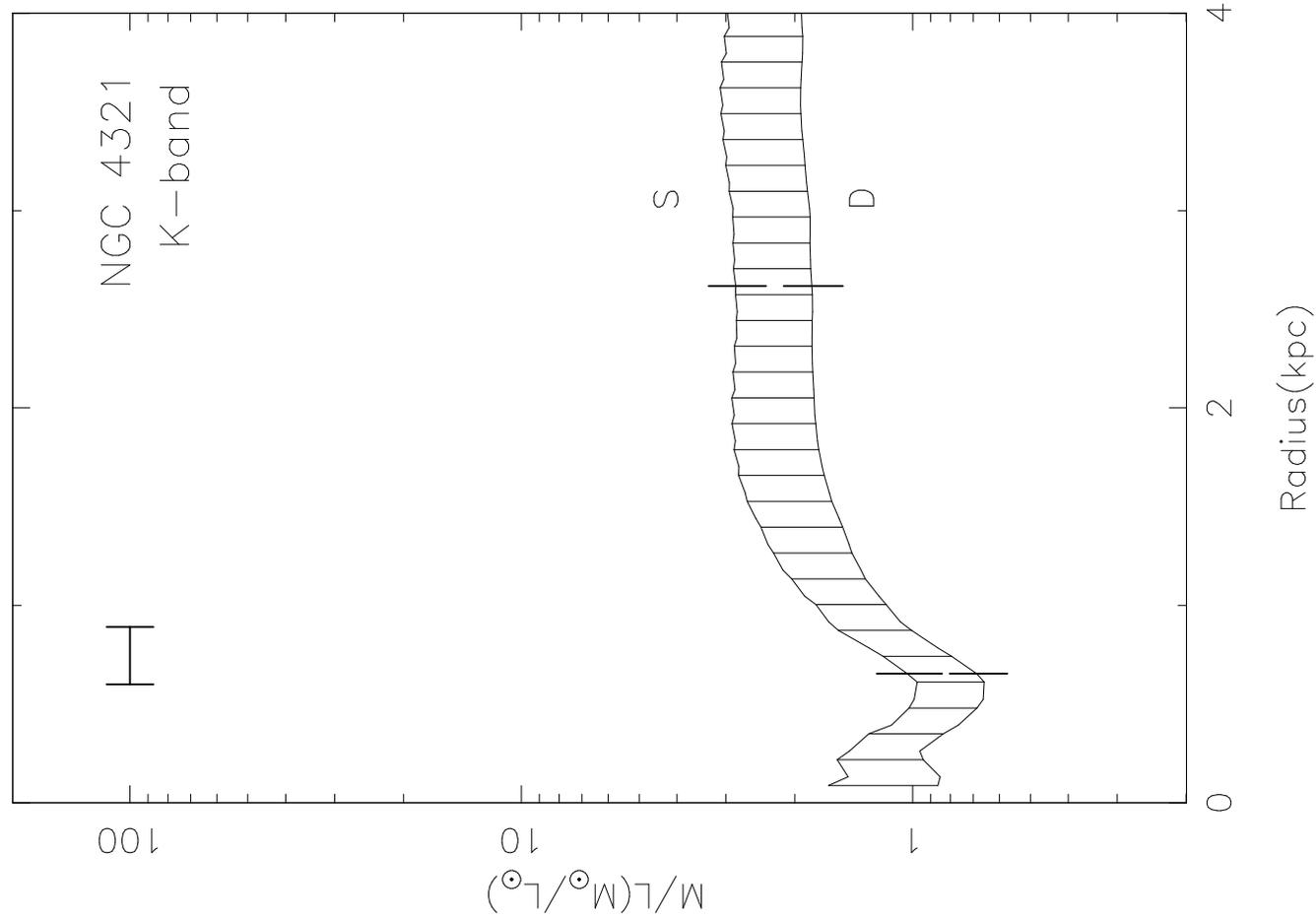
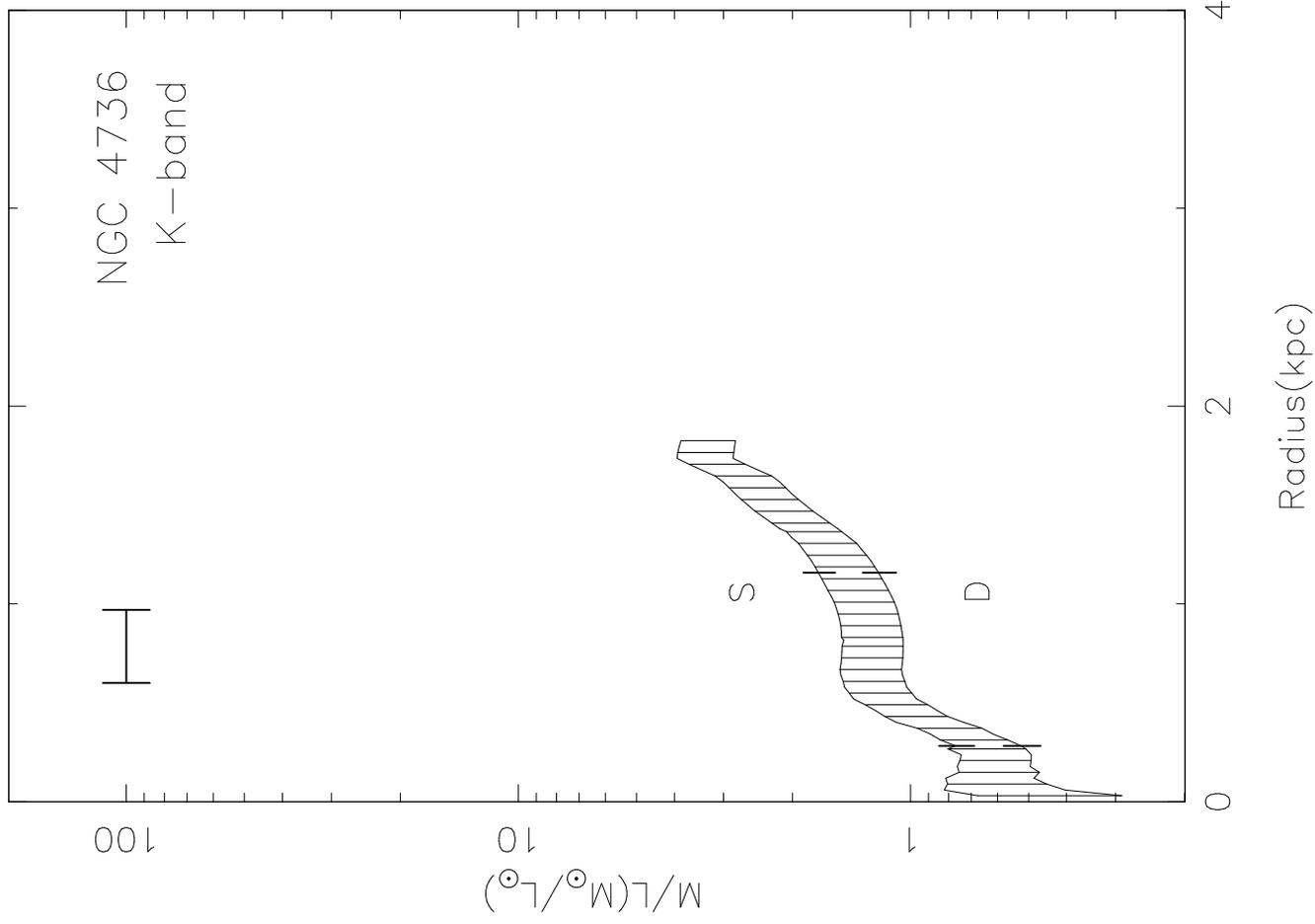
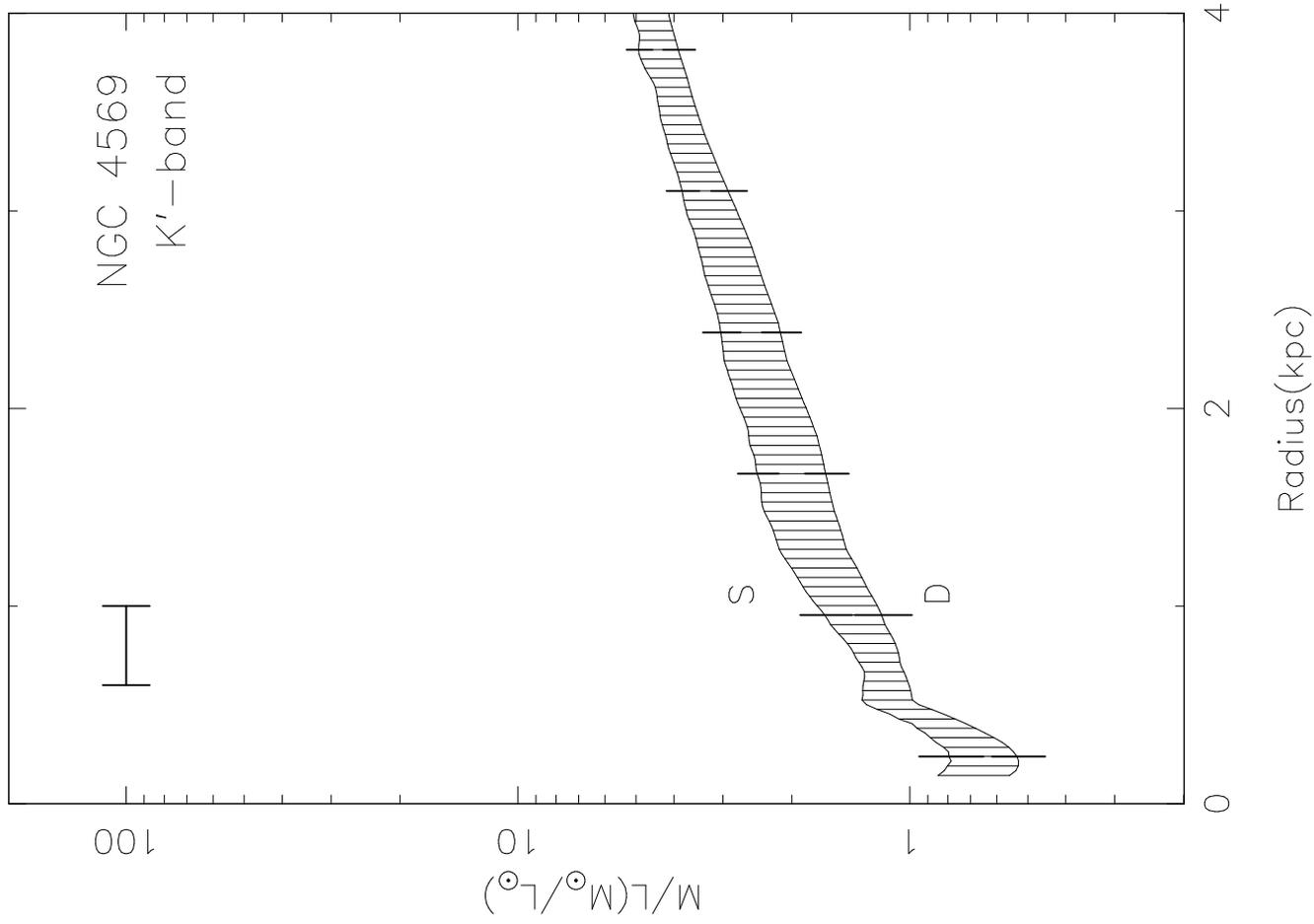
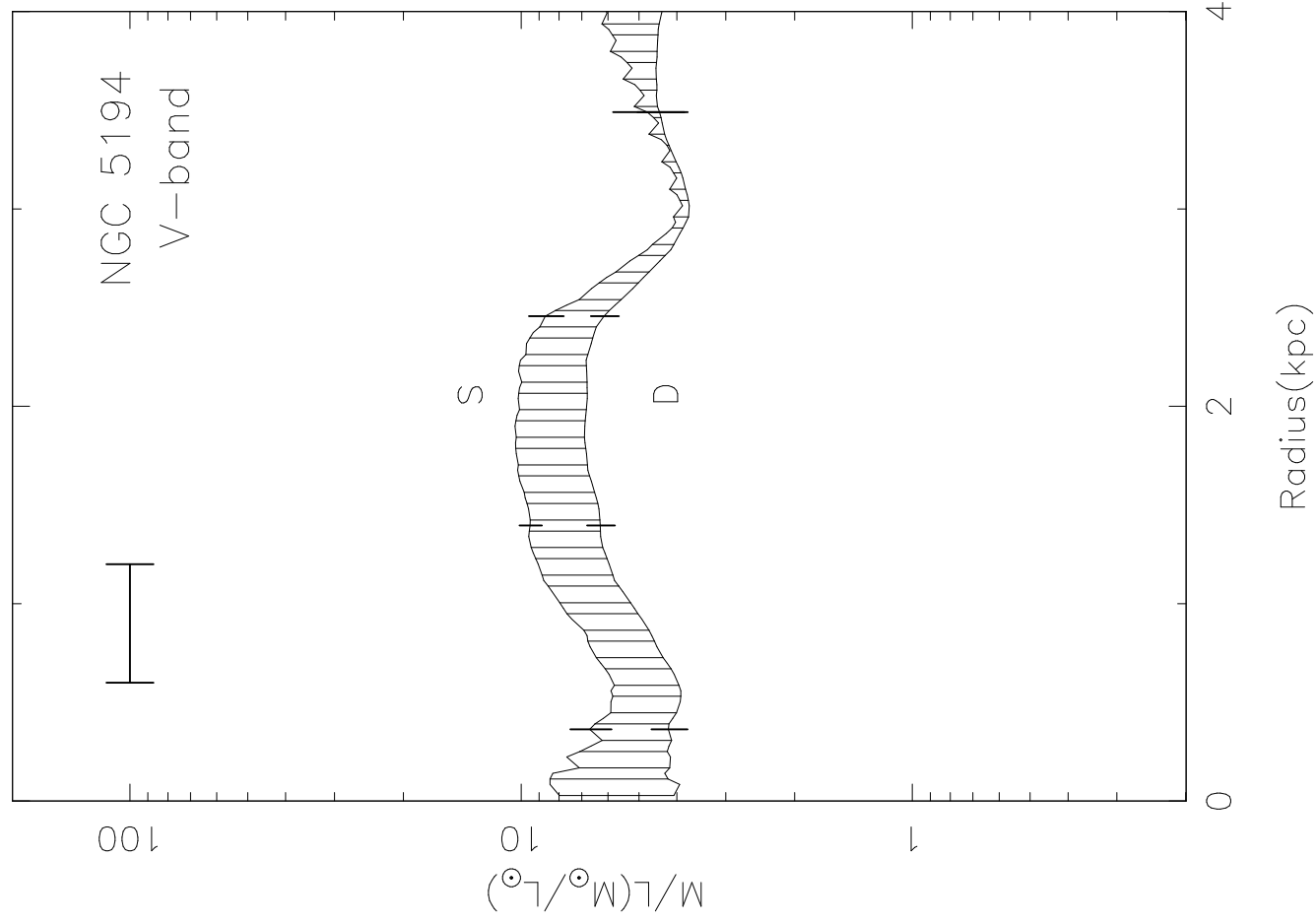
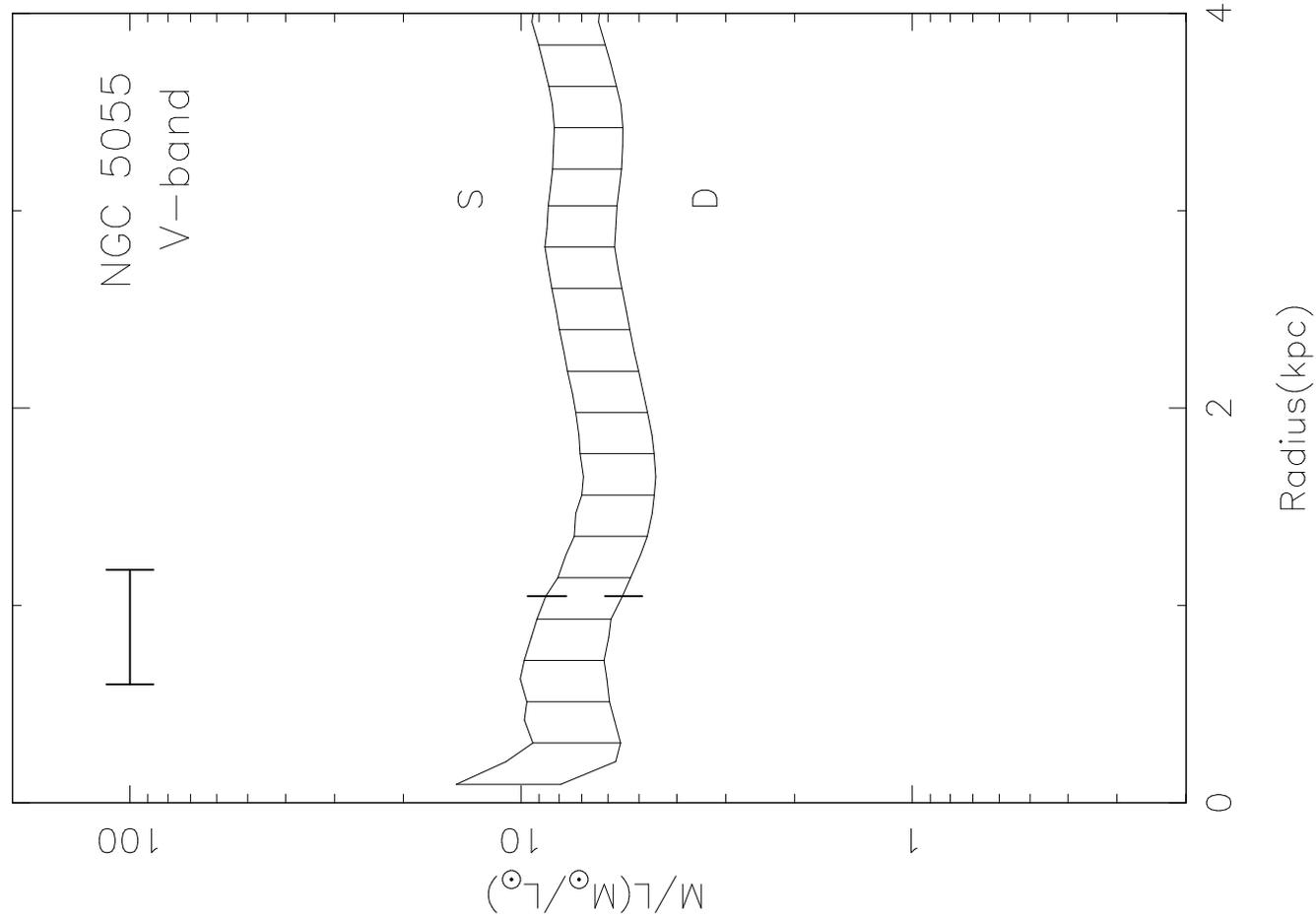
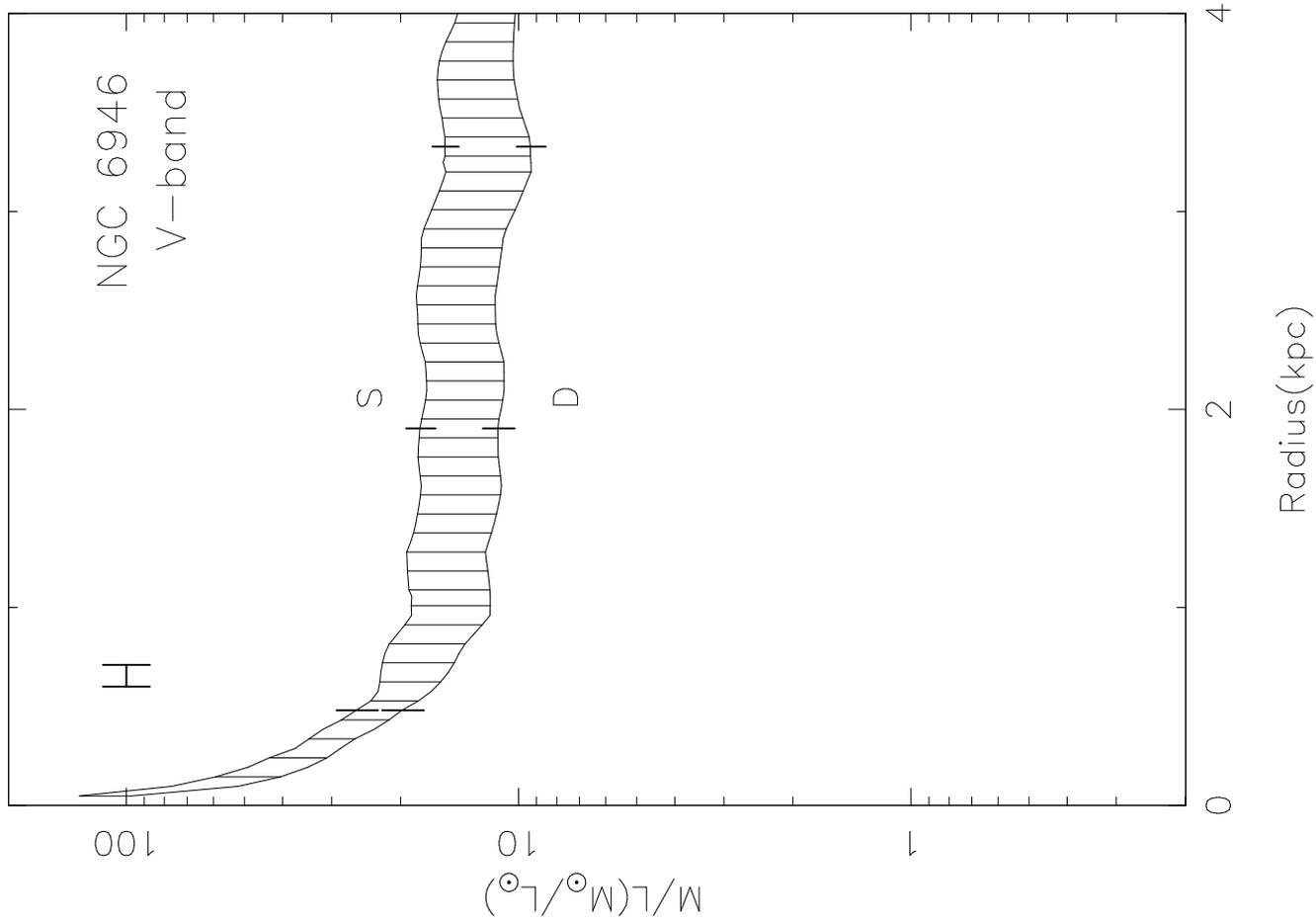
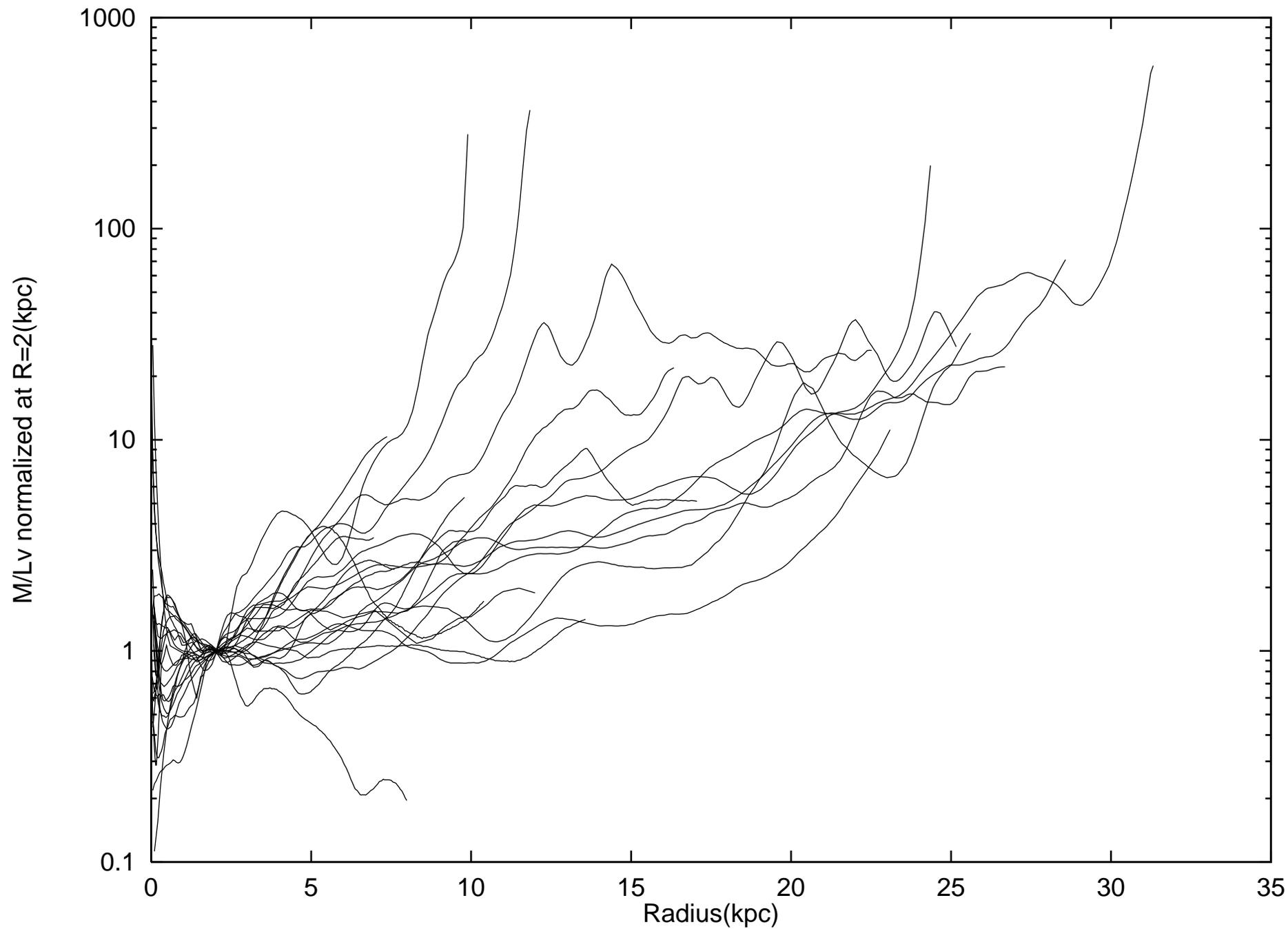
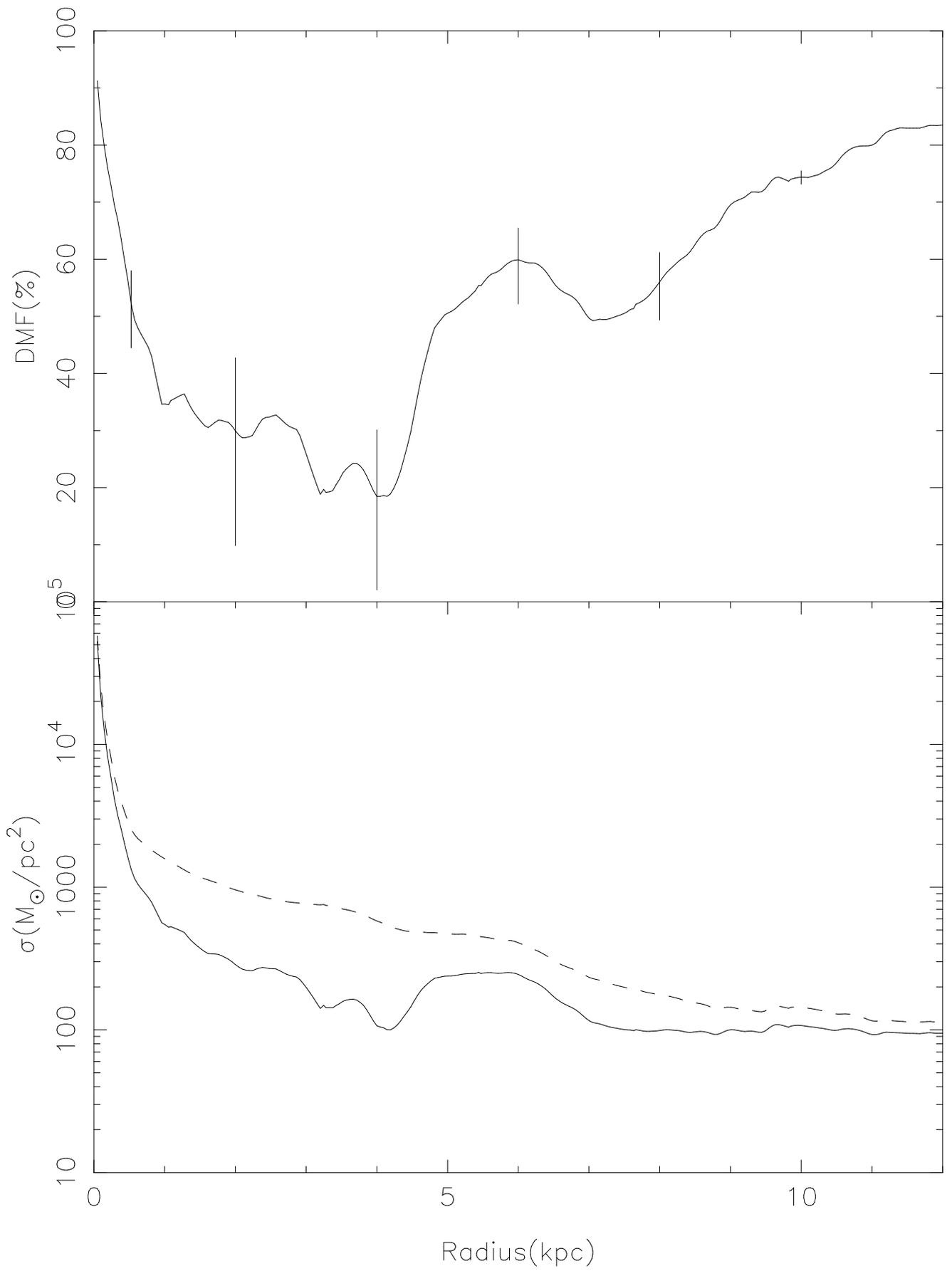
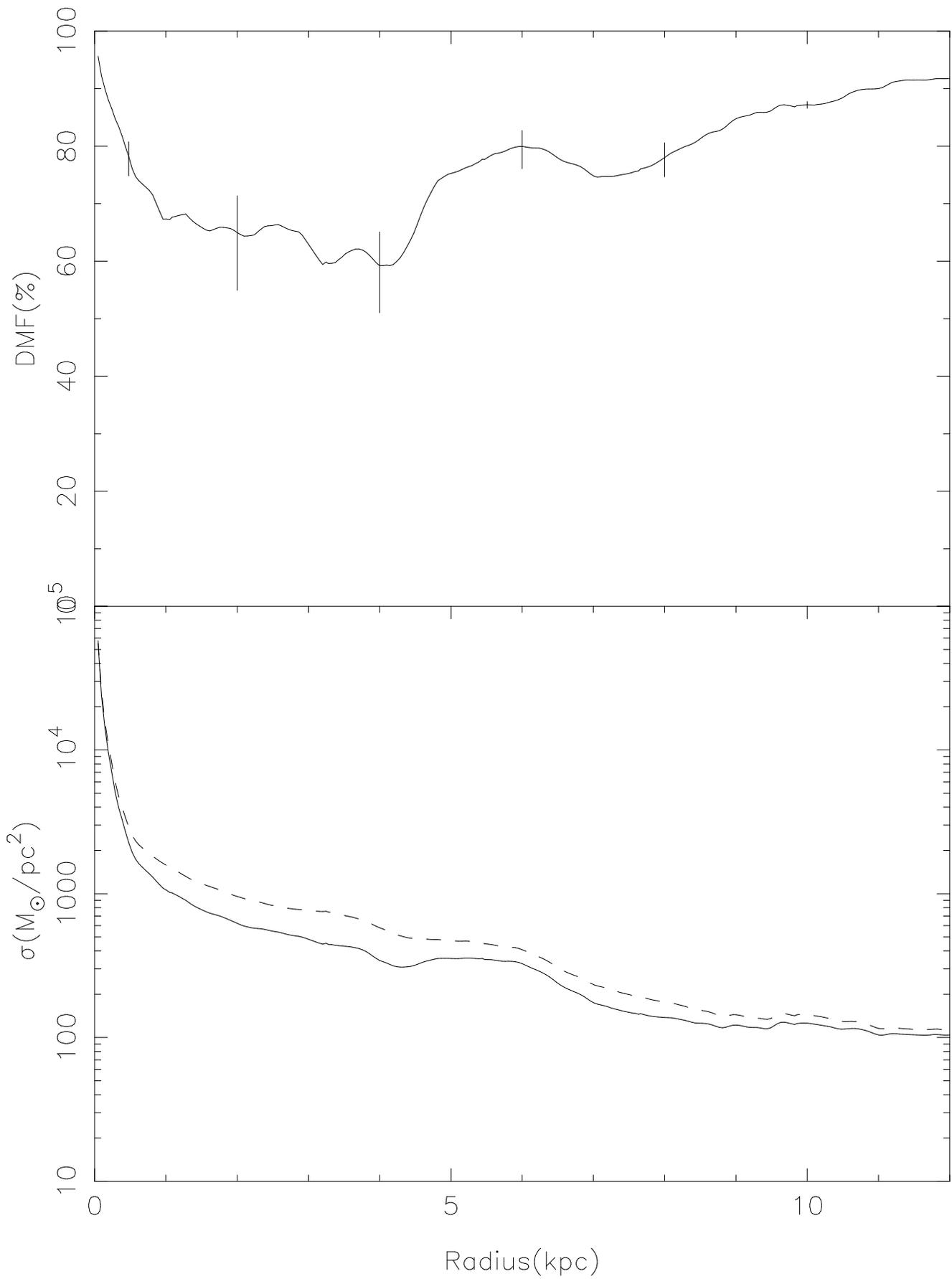
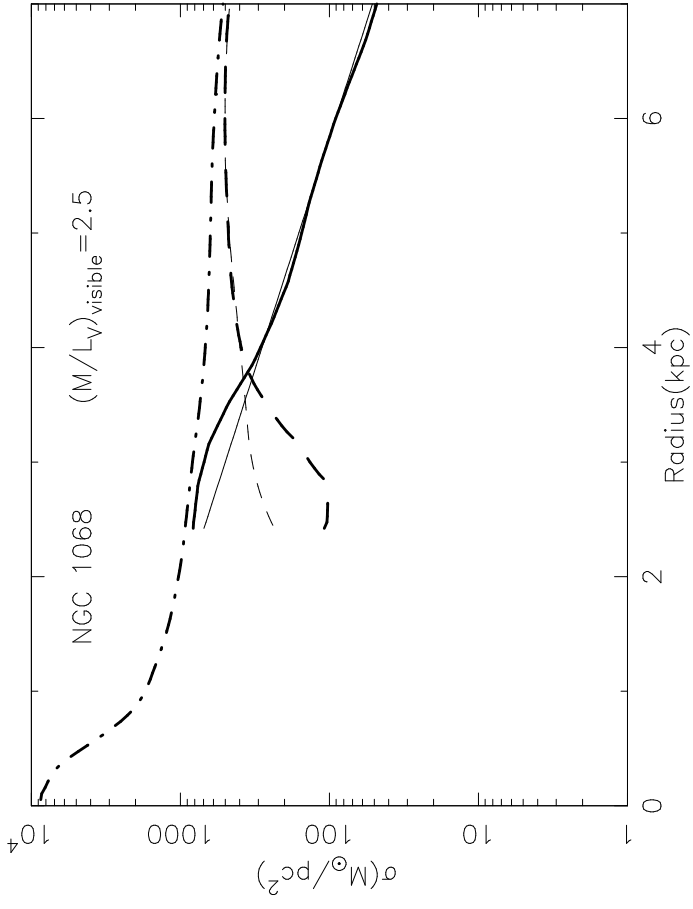
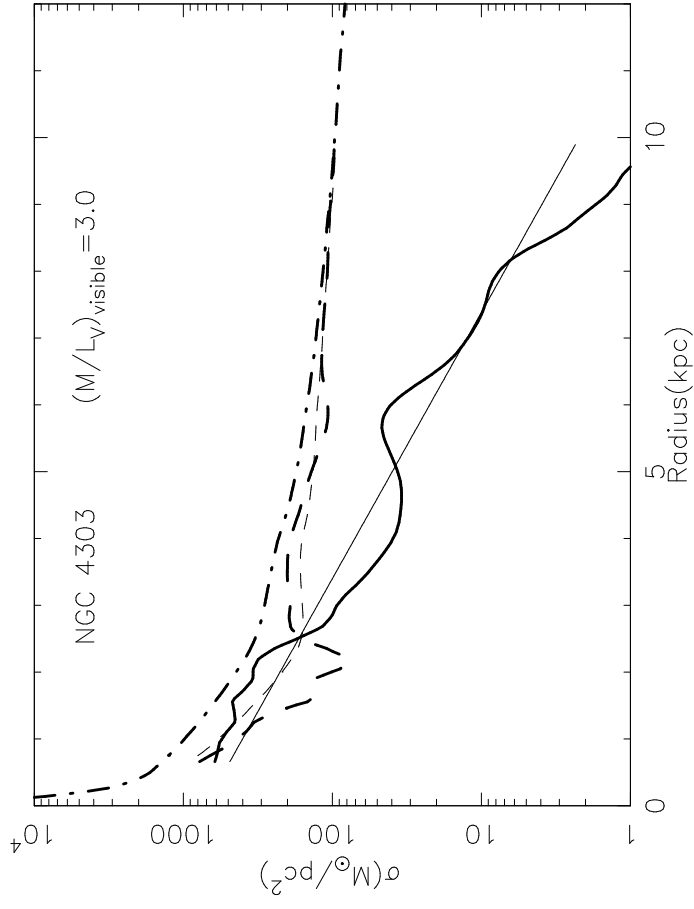
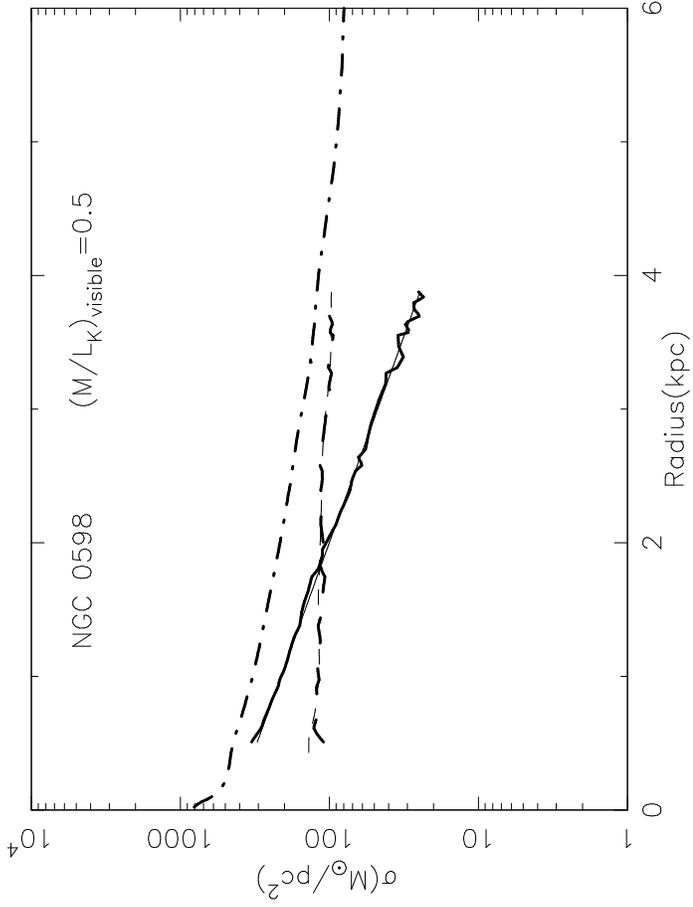
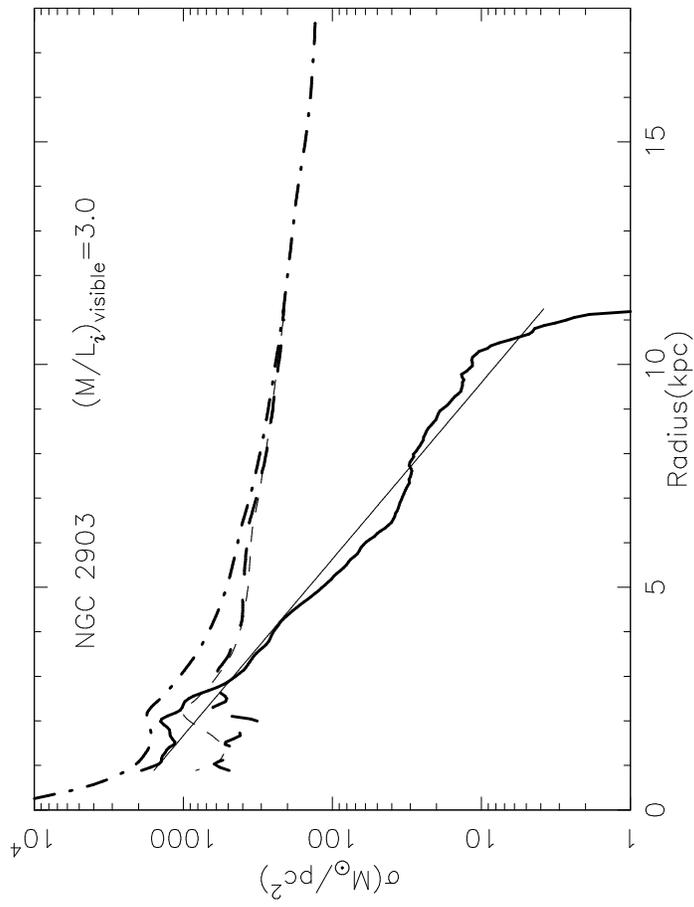
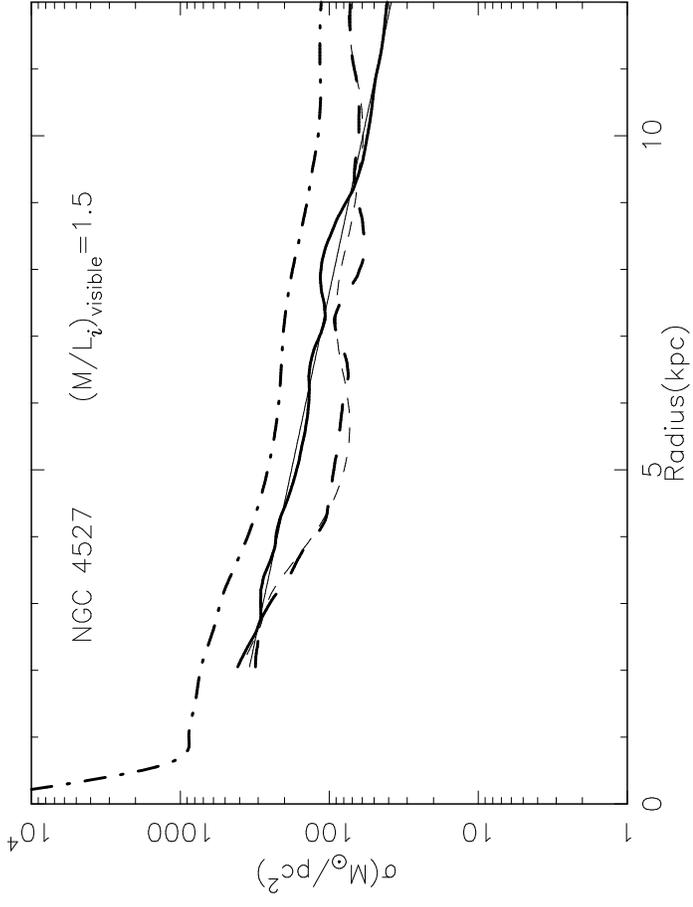
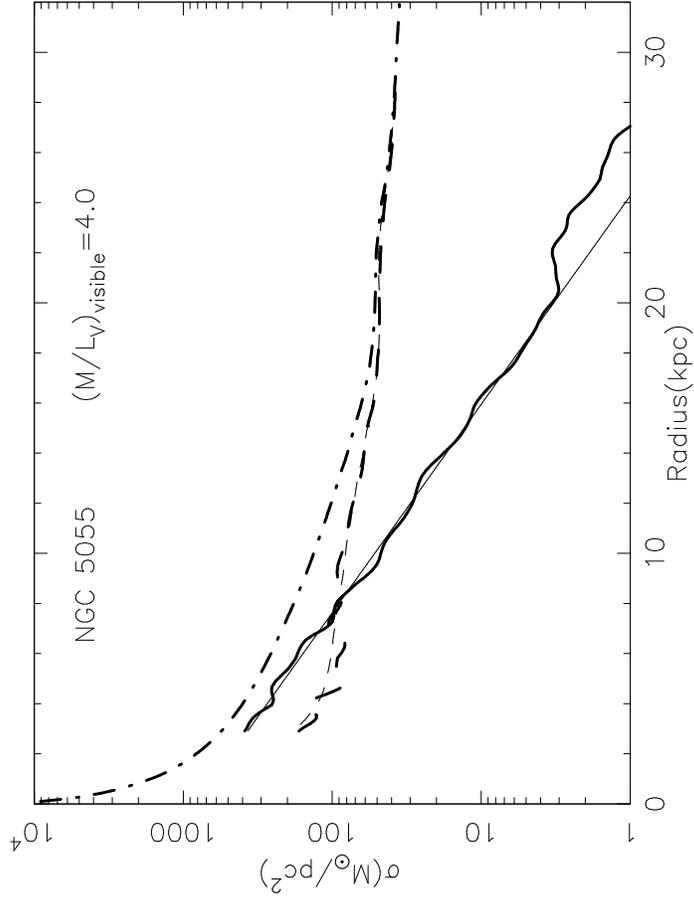
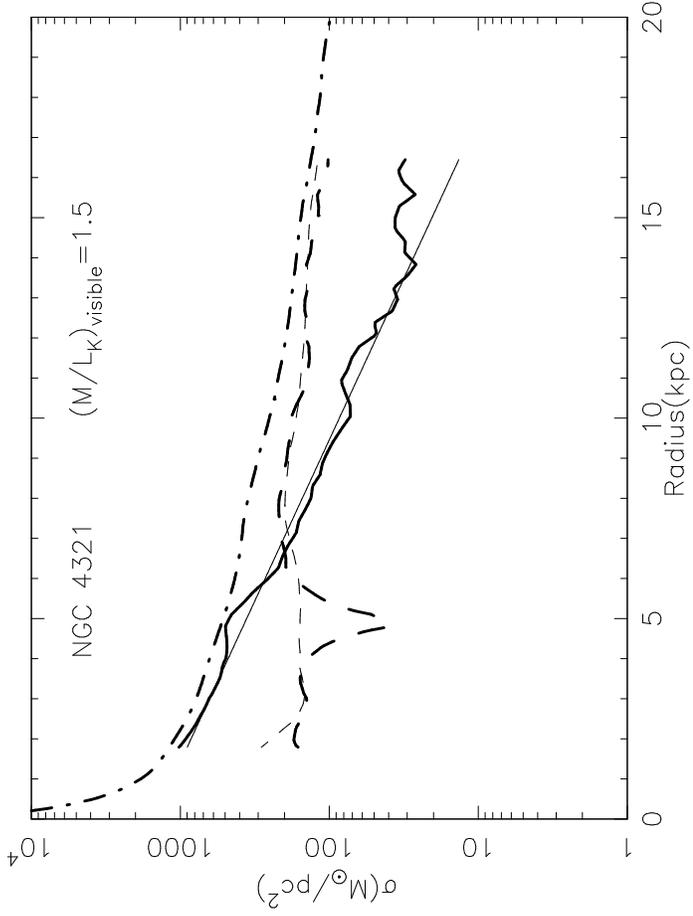
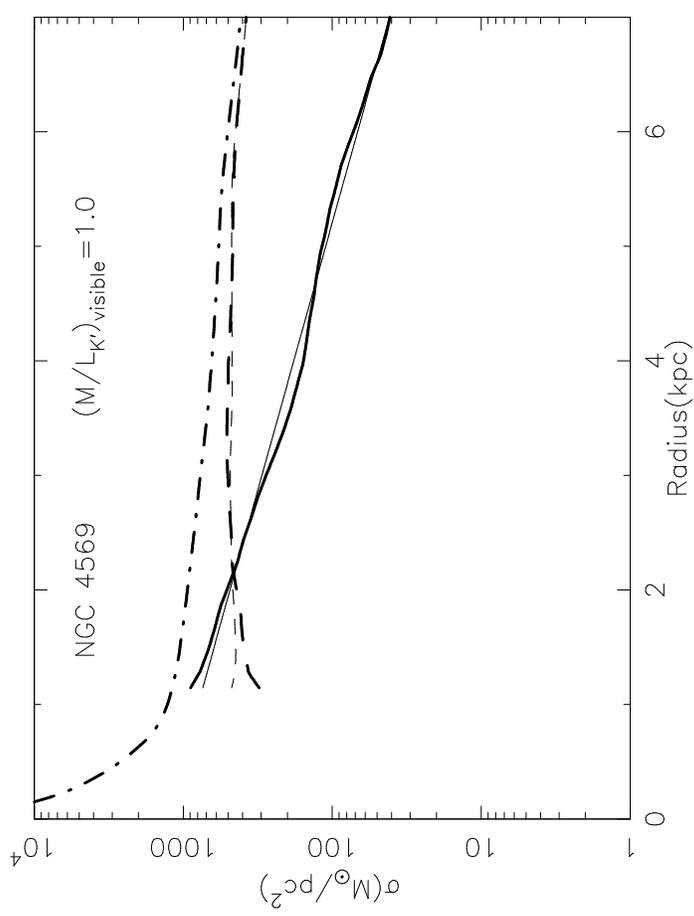
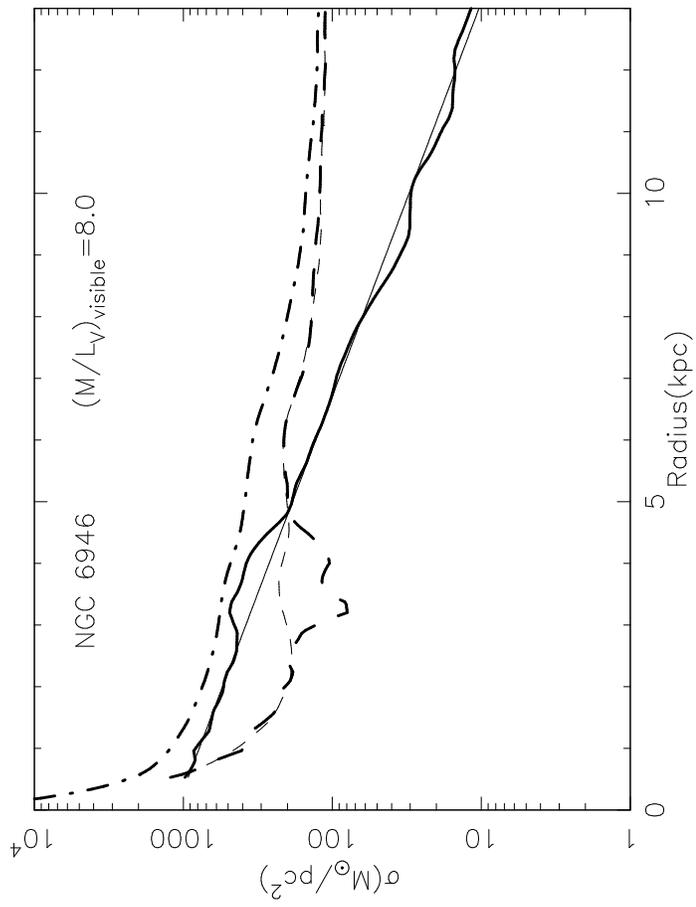
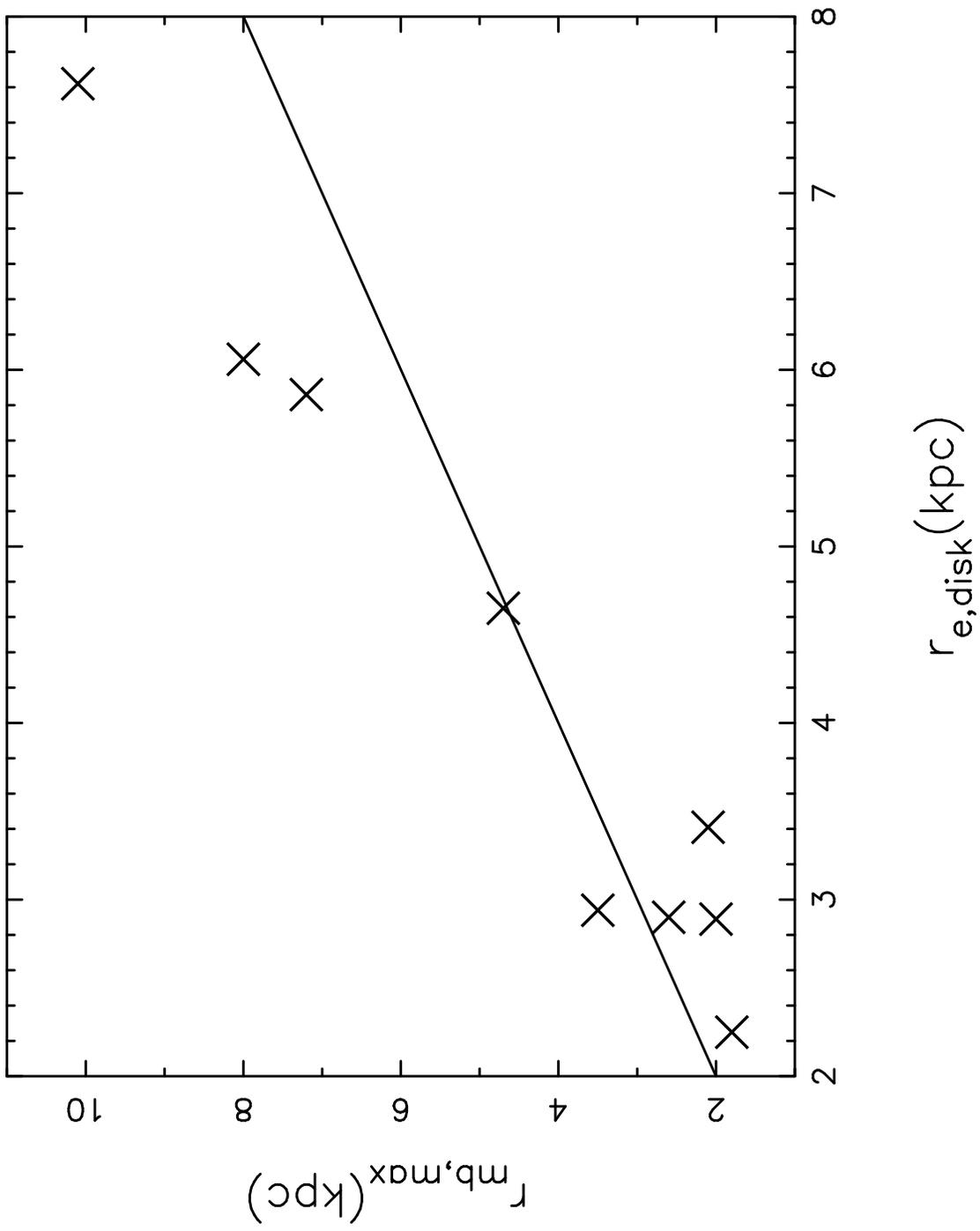
\end{document}